\documentclass[onecolumn,fleqn,usenatbib]{mnras}

\usepackage{newtxtext,newtxmath}

\usepackage[T1]{fontenc}
\usepackage{ae,aecompl}

\usepackage{graphicx}   
\usepackage{amsmath}    

\usepackage{threeparttable} 
\usepackage{threeparttablex} 
\usepackage{longtable,lscape}
\usepackage{rotating}

\defcitealias{burstein82}{BH82}
\defcitealias{burns18}{B18}
\defcitealias{folatelli10}{F10}
\defcitealias{freedman19}{F19}
\defcitealias{hamuy93a}{H93a}
\defcitealias{hamuy93b}{H93b}
\defcitealias{hamuy96a}{H96a}
\defcitealias{hamuy96b}{H96}
\defcitealias{hamuy96c}{H96c}
\defcitealias{kattner12}{K12}
\defcitealias{phillips99}{P99}
\defcitealias{riess16}{R16}
\defcitealias{riess19}{R19}
\defcitealias{schlafly11}{SF11}
\defcitealias{schlegel98}{SFD98}

\title[The value of the Hubble-Lema\^itre constant]{The value of the Hubble-Lema\^itre constant queried by Type Ia Supernovae: A journey from the Calán-Tololo Project to the Carnegie Supernova Program}
\author[Hamuy {\it et al.}]{Mario Hamuy$^{1,2}$\thanks{mhamuy@aura-astronomy.org}, R\'egis Cartier$^{3}$ Carlos Contreras$^{4}$ and Nicholas B. Suntzeff $^{5,6}$\\ 
$^{1}$Vice President and Head of Mission of AURA-O in Chile, Avda Presidente Riesco 5335 Suite 507, Santiago, Chile \\
$^{2}$Hagler Institute for Advanced Studies, Texas A\&M University, Texas, USA\\
$^{3}$Cerro Tololo Inter-American Observatory, NSF's National Optical-Infrared Astronomy Research Laboratory, Casilla 603, La Serena, Chile\\
$^{4}$Carnegie Observatories, Las Campanas Observatory, Casilla 601, La Serena, Chile \\
$^{5}$George P. and Cynthia Woods Mitchell Institute for Fundamental Physics and Astronomy,  College Station, TX 77843\\
$^{6}$Department of Physics and Astronomy, Texas A\&M University, College Station, TX 77843 \\
}

\begin{document}

\newcommand{\vdag}{(v)^\dagger}
\newcommand\aastex{AAS\TeX}
\newcommand\latex{La\TeX}
\newcommand\Ho{$H_0$}
\newcommand\dm{$\Delta m_{15}(B)$}
\newcommand{\kms}{\hbox{$ \, \rm km\, s^{-1}$}}
\newcommand{\kmsmpc}{\hbox{$ \, \rm km\, s^{-1} \, Mpc^{-1}$}}
\newcommand{\None}[1]{}

\newcommand\plottwo[2]{{%
\typeout{Plottwo included the files #1 #2}
 \centering
 \leavevmode
 \columnwidth=.45\columnwidth
 \includegraphics[width={\eps@scaling\columnwidth}]{#1}%
 \hfil
 \includegraphics[width={\eps@scaling\columnwidth}]{#2}%
}}%

\pagerange{\pageref{firstpage}--\pageref{lastpage}} \pubyear{2020}

\maketitle

\label{firstpage}

\begin{abstract}

We assess the robustness of the two highest rungs of the “cosmic distance ladder” for Type Ia supernovae and the determination of the Hubble-Lema\^itre constant. In this analysis, we hold fixed Rung 1 as the distance to the LMC determined to 1\% using Detached Eclipsing Binary stars. For Rung 2 we analyze two methods, the TRGB and Cepheid distances for the luminosity calibration of Type Ia supernovae in nearby galaxies. For Rung 3 we analyze various modern digital supernova samples in the Hubble flow, such as the Calán-Tololo, CfA, CSP, and Supercal datasets. This metadata analysis demonstrates that the TRGB calibration yields smaller \Ho\ values than the Cepheid calibration, a direct consequence of the systematic difference in the distance moduli calibrated from these two methods. Selecting the three most independent possible methodologies/bandpasses ($B$, $V$, $J$), we obtain \Ho=69.9$\pm$0.8 and \Ho=73.5$\pm$0.7 \kmsmpc from the TRGB and Cepheid calibrations, respectively. Adding in quadrature the systematic uncertainty in the TRGB and Cepheid methods of 1.1 and 1.0 \kmsmpc, respectively, this subset reveals a significant 2.0 $\sigma$ systematic difference in the calibration of Rung 2. If Rung 1 and Rung 2 are held fixed, the different formalisms developed for standardizing the supernova peak magnitudes yield consistent results, with a standard deviation of 1.5 \kmsmpc, that is, Type Ia supernovae are able to anchor Rung 3 with 2\% precision. This study demonstrates that Type Ia supernovae have provided a remarkably robust calibration of R3 for over 25 years.

\end{abstract}


\section{Introduction}

After one century of research, the advances of recent years both in the field of theory
and experimentation have allowed us to witness remarkable progress in our understanding
of the Universe on large scales. A concordant $\Lambda$CDM cosmological model is able to
reproduce the evolution of the Universe from the epoch of recombination, characterized by the remnants effects of
density fluctuations of quantum origin, to its complex current large scale structure.
Such a model is geometrically flat, composed of cold dark matter, and has a dominant
component of dark energy that is responsible for the current acceleration of the Universe. 
Remarkably, one requires only six cosmological parameters to define the basic cosmology as has
been observationally demonstrated by the WMAP and Planck missions.

\smallskip
Within the $\Lambda$CDM model, the Hubble-Lema\^itre constant (\Ho) is arguably the most important
cosmological parameter. By definition it corresponds to the expansion rate of the Universe
at the present time. It sets the size, age, and critical density of the Universe, and pervades
virtually all models in extra-galactic research. Ever since the discovery of the cosmic
expansion in 1927-29 \citep{Lemaitre27,hubble29}, there has been a continuous effort from
the astronomical community to measure its value, with the range of experimentally measured
values of \Ho\ decreasing over time, from $\sim$500 \kmsmpc\ to a narrow interval of only 67-74 \kmsmpc.

\smallskip
The traditional method has consisted in measuring luminosity distances to galaxies in the smooth
Hubble flow from bright astronomical sources with properly calibrated luminosities. This
approach has required the calibration of a series of increasingly brighter astrophysical sources,
which, altogether is known as the ``cosmic distance ladder'' (CDL). Many different techniques 
have been attempted in order to build the ladder and determine the value of \Ho. 
In this work we will focus in a particularly successful architecture, which is based on enormous improvements 
over the past three decades in: (1) improving our ability to measure precise (5-7\%) distances to individual Type Ia supernovae (SNe~Ia), (2) establishing Cepheid or Tip of the Red Giant Branch (TRGB) distances, with the Hubble Space Telescope (HST), to a growing sample of galaxies having hosted SNe~Ia, and (3) improving the determination of the distance to the Large Magellanic Cloud (LMC) or other very nearby galaxies.
The concatenation of these techniques creates a three-rung ladder where all links are essential for the purpose of determining the value of \Ho\ and {\it none is less important than the other}.
Several authors claim today that this concatenation of methods can lead to a $\sim$1\%~precision in
the measurement of \Ho\ \citep[R16, B18, R19 \& F19]{riess16,burns18,riess19,freedman19}. However,
the reported values range between 74.22$\pm$1.82 \citepalias{riess19} and 69.8$\pm$0.8 \kmsmpc
\citepalias{freedman19}, which shows that the 1\% precision is a future goal for the CDL.

\smallskip
Other experimental approaches sensitive to \Ho\ but independent of the cosmic ladder in the local Universe
have been advocated in recent years such as the measurement of temperature anisotropies in the cosmic
microwave background (CMB). The exceptional data provided by the WMAP and Planck satellites have allowed
precise determinations of the Hubble-Lema\^itre constant using the CMB data alone, namely \Ho=70.0$\pm$2.0 and
\Ho=67.4$\pm$0.5 \kmsmpc, respectively \citep{hinshaw13,planck1}. It must be kept in mind that these values 
are {\it indirect} constraints based on a flat $\Lambda$CDM cosmological model.

\smallskip
The measurement of the angular diameter of the baryon acoustic oscillation (BAO) feature is also
sensitive to the expansion history. However, the BAO depends on the sound horizon measured by the CMB, 
so the two \Ho\ results are not independent. The parameters yielded by the BAO experiment using the Sloan Digital Sky Survey III
data anchored to the Planck CMB data leads to \Ho=67.6$\pm$0.5 \kmsmpc \citep{alam17}, thus providing further evidence for the
six parameter cosmological model fit by the Planck collaboration. As shown by \citet{addison18}, when Ly$\alpha$ BAO data are combined with CMB data, the solutions for \Ho\ obtained from WMAP and Planck agree even better and with smaller uncertainties, namely, \Ho=68.3$\pm$0.7 and \Ho=68.1$\pm$0.6 \kmsmpc, respectively.

\smallskip
Strong gravitational lenses afford another route for the cosmic ladder, yet model-dependent.
This method consists in measuring time delays between different images of a background quasar lensed by a foreground
galaxy and modeling the lens mass distribution. Recently, \citet{wong19} presented a measurement of the Hubble-Lema\^itre constant of
73.3$\pm$1.8 \kmsmpc\ from six lens systems.

\smallskip

The Megamaser Cosmology Project has recently obtained another measurement of the Hubble-Lema\^itre constant independent from the cosmic ladder. Their analysis yielded distances from six magamaser-hosting galaxies, which led to a constraint to the Hubble-Lema\^itre constant of \Ho=73.9$\pm$3.0 \kmsmpc\
\citep{reid19}.

\smallskip
The value of the Hubble-Lema\^itre constant is a long standing controversy. Thirty years ago the debate was between values of 50 and 100 \kmsmpc. Since then we have seen a notable progress but, as the precision of our measurements has increased, we find ourselves once again with two camps advocating significant, although small, differences between 67-74 \kmsmpc. This $\sim$ 10\% difference is 5$\sigma$ beyond their internal uncertainties, a difference too large for the precision astronomy era, that could be explained for our current inability to identify and handling the systematics in the distance ladder, or for the lack of a complete understanding of the early Universe physics or to later variations in the behaviour of dark energy. The latter makes the \Ho\ problem even more interesting to solve.

\smallskip
The purpose of this paper is to make a thorough revision of the setting of the CDL which, as shown above, is currently
delivering internally discrepant values between 74.22$\pm$1.82 \citepalias{riess19} and 69.8$\pm$0.8 \kmsmpc\
\citepalias{freedman19}. Our goal is to focus the attention into the heart of the distance ladder method, that is, we
will not discuss the Rung 1 (the determination of the LMC distance), which we assume well determined, but we will
reanalyze in detail the second and third rungs of the distance ladder. For the Rung 2 we will investigate the
impact on the value of \Ho\ using recent Cepheid and TRGB relative distances anchored to the LMC. For Rung 3 we will employ different samples of SNe~Ia, starting with the first set of digital light curves obtained in the
early 90s, combined with different methodologies to standardize their luminosities, which will allow us to assess the
consistency and the systematics uncertainties of the SNe~Ia technique.

\smallskip
This paper is organized as follows. In section \ref{cdl} we review the latest advances in the establishment of the
cosmic distance ladder. In section \ref{analysis} we analyze the systematics of the CDL. First, we focus on Rung 2 assessing the implications of adopting the Cepheid and TRGB distances for the calibrations of several SN~Ia samples. Then we assess the robustness of Rung 3 employing 30 combinations of SN~Ia samples observed in optical and near-infrared (NIR) bandpasses, and six different methodologies for the standardization of the SN peak luminosities. Finally, in section \ref{conclusions} we summarize the main conclusions of this paper.

\section{The Cosmic Distance Ladder} \label{cdl}

The traditional method to measure \Ho\ consists in establishing a cosmic distance ladder whose highest third rung provides a direct measurement of the cosmic expansion rate from galaxies in the smooth Hubble flow\footnote{Defined as the limit where the distance modulus errors are roughly matched to the peculiar velocity. For a peculiar velocity of 200 \kms\ and a 0.1 mag error in distance modulus, the limit of the smooth Hubble flow is $z$=0.014.}. This last step has been approached using various types of objects, but the most precise methods remain those involving SNe~Ia \citep{freedman01}. Thus, the modern determination of \Ho\ involves the following three steps (or rungs), namely, (1) the measurement of the distance to a nearby galaxy such as the LMC, NGC 4258, M31 or parallaxes in the Milky Way; (2) distance determinations to other nearby SN Ia host galaxies (distance modulus $\mu$ $<$ 33), relative to the first anchor, via the traditional Cepheid method or the most recent TRGB technique; and (3) the measurement of distances to SNe~Ia in the Hubble flow (range of redshifts $z$=0.01-0.1, or $\mu$=33-38), applying the inverse square law to their apparent magnitudes and their intrinsic luminosities calibrated via Cepheids or TRGB stars.

\smallskip
{\it The First Rung.} Rung 1 (R1) has been established with four methods: (1) the modelling of masers in the galaxy NGC~4258 which yields a distance modulus of 29.40$\pm$0.02 \citep{reid19}; (2) Detached Eclipsing Binary stars (DEBs) which yield a distance modulus for the LMC of 18.48$\pm$0.02 \citep[precision of 1\% in distance;][]{pietrzynski19}; (3) trigonometric parallaxes of Milky Way Cepheids \citep{vanleeuwen07,benedict07} and; (4) DEBs in M31 \citep{ribas05,vilardell10}.

\smallskip
The calibration of Rung 2 has been anchored to one or more of these four calibrations. For instance, \citetalias{riess16} adopted all four of these calibrations for the determination of the Cepheid luminosities in 19 galaxies which have hosted SNe Ia. Instead, \citetalias{freedman19} adopted solely the LMC distance calibration measured from DEBs for the measurement of the TRGB in 18 SNe~Ia host galaxies. This complicates a direct comparison of both methods and their associated systematic uncertainties.

\smallskip
\citetalias{riess19} updated the Cepheid calibration to be anchored solely to the LMC DEB distance {\it but did not provide the revised individual galaxy distances}. They found that the net effect of changing the zero point (R1) of the CDL is an increase in \Ho\ from 73.24 to 74.22 \kmsmpc, 1.34\% relative to their 2016 value. This increase in the value of \Ho\ can be translated into a global correction of -0.029 magnitudes to their 2016 distance moduli catalog. This allows us to establish a common ground for the first rung, leaving the R1 calibration out of this discussion, to focus on the assessment of the R2 calibration using Cepheid and TRGB techniques, and the R3 calibration using SNe~Ia.

\smallskip
{\it The Second Rung.} The calibration of Rung 2 (R2) has been historically approached using the classical Cepheid Leavitt law (the P-L relation), ever since the pioneering work of Sandage et al. and Freedman et al. in the 1990 decade using the Hubble Space Telescope (HST). This approach has been improved significantly as the sensitivity of HST has allowed the discovery and characterization of Cepheid stars in a greater and more distant sample of SN Ia host galaxies and through the addition of new SNe Ia that have exploded in nearby galaxies over recent years. There are 19 SNe~Ia possessing Cepheid-calibrated distances in the range $\mu$=29.1-32.9 (\citetalias{riess16}, \citetalias{riess19}). 
\smallskip
In the last years an alternative technique has matured allowing the determination of precise distances to nearby galaxies by identifying the locus of the TRGB stars in a colour magnitude diagram \citep{lee93,beaton16}. The TRGB technique affords a competitive and independent method for the calibration of R2. This work has been vigorously championed by the Carnegie-Chicago Hubble Project (CCHP) in a series of eight papers that introduce the method, measure distances to 13 SNe~Ia host galaxies, bring to a common scale five additional galaxies measured by \citet{jang15,jang17}, thus raising to 18 the total number of SNe~Ia host galaxies with TRGB distances \citepalias{freedman19}.

\smallskip
The Cepheid and TRGB sets of distance calibrators overlap in ten galaxies, thus allowing a measurement of the systematic differences in the calibration of the SNe~Ia luminosities and of R2, and the implications in the determination of the value of \Ho, as the reader will see in section \ref{rung2}.

\smallskip
{\it The Third Rung.} The third and highest rung of the cosmic distance ladder (R3) has been reached with different methods such as galaxies themselves using the Tully-Fisher \citep{giovanelli97}, Surface Brightness Fluctuations \citep{tonry97}, Planetary Nebulae \citep{feldmeier07}, and Faber-Jackson \citep{faber76} techniques. However, it has been demonstrated that supernovae play the most competitive role in this endeavour. Several approaches have been developed for Type II SNe such as 
the Standardized Candle Method \citep[SCM;][]{olivares10}, the Photometric Colour Method \citep[PCM;][]{dejaeger15,dejaeger20}, and the Photospheric Magnitude Method \citep[PMM;][]{rodriguez14}. Undoubtedly, the most precise approach to measure the local expansion rate of the Universe has been achieved using SNe~Ia, thanks to their enormous brightness and the standardization of their peak luminosities to reach an unrivaled level of 5-7\% precision in distance.

\smallskip
The success of SNe~Ia as precise distance indicators goes back to the pioneering work of \citet{kowal68} using photographic photometry. The potential of supernovae was also emphasized by \citet{sandage70} in his famous paper ``Cosmology: A search for two numbers", which was later confirmed using high-precision digital photometry. The first large, multiband CCD sample of distant SNe was produced by the Calán-Tololo survey carried out from the Cerro Tololo Inter-American Observatory between 1989-1993 \citep{hamuy93a} which ended with the publication of 29 $BVRI$ optical SNe Ia light curves \citep{hamuy96c}. In 1993 astronomers of the Center for Astrophysics (CfA) started a photometric monitoring campaign of SNe Ia using CCD detectors at the Fred Lawrence Whipple Observatory, which yielded a first release of 22 SNe Ia optical ($BVRI$) light curves \citep{riess99}, a second release of 44 SNe Ia \citep{jha06} and more recently a third release (CfA 3) of 185 SNe Ia observed between 2001-2008 \citep{hicken09}. The extensive Lick Observatory Supernova Search (LOSS) program carried out since 1998 has produced more than 200 $BVRI$ SNe~Ia light curves \citep{li00,filippenko01,ganeshalingam10,stahl19}. The Carnegie Supernova Program (CSP) carried out between 2004-2009 from Las Campanas Observatory (LCO) meant a significant advance in the quality of the SN Ia optical light curves, thanks to the use of a uniform photometric system, in situ measurements of the full transmission curves for the telescope/filter/CCD system, and by expanding the survey to NIR $YJHK$ bands \citep{hamuy06}. The CSP released two initial datasets \citep{contreras10,strizinger11}, and a third final data release published by \citet{krisciunas17} which contains the overall CSP I dataset of 134 SNe~Ia. Since 2017, the Foundation Supernova Survey has been obtaining $griz$ light curves with the Pan-STARRS telescope on the peak of Haleakala on the island of Maui. A first data release of 225 SNe~Ia was recently published by \citet{foley18} \footnote{We omit from this summary the high-z surveys designed for the measurement of dark energy.}.

\smallskip
Along with obtaining larger samples of SNe~Ia with increasing precision, observing cadence and wavelength coverage, the success of the SNe~Ia method has critically relied on the developments of novel techniques for the standardization of their peak luminosities, such as the correction light curve decline rate \citep{phillips93,riess95}, host-galaxy extinction \citep{riess96,phillips99}, and, most recently, host-galaxy mass \citep{kelly10,sullivan10,burns18}. The great variety of distant SN Ia samples and the different techniques implemented in the standardization of their peak magnitudes afford an opportunity to study possible systematic differences in the SNe~Ia method, as will be seen in the following section.

\section {Systematics in the value of the Hubble-Lema\^itre constant from the cosmic distance ladder} \label{analysis}

The purpose of this section is to derive an additional evaluation of the systematic uncertainties in the determination
of \Ho\ from the CDL approach to that addressed by \citet{freedman19} and \citet{riess19}. As mentioned in the previous section, our strategy consists in leaving the R1 calibration out of this discussion, to focus on the assessment of R2 applying both the Cepheid and the TRGB calibrations to several modern samples of nearby SNe~Ia, and study the R3 calibration using various datasets and methodologies for standardizing the luminosities of distant SNe~Ia. Having multiple datasets/methodologies affords a novel opportunity to empirically assess the internal consistency, possible systematic differences in the SNe~Ia technique, and derive a more precise value of \Ho\ by combining independent datasets.

\smallskip
Ever since the work of \citet{rust74}, \citet{pskovskii77} and \citet{phillips93}, it was unambiguously demonstrated that SNe~Ia were
not perfect standard candles in the optical bands, and that their peak magnitudes were correlated with the width of
their light curves. The gathering of the first dataset of digital photometry for SNe in the Hubble flow by the
Calán-Tololo survey confirmed such correlation and proved that it was possible to successfully standardize their
peak magnitudes to unrivaled levels of 0.14 mag, or 7\% in distance \citep{hamuy95,hamuy96b}. As the
distant samples became more numerous, it was possible to identify additional parameters such as SN colour
\citep{lira96,riess96,tripp98,phillips99}, or host galaxy properties to further standardize the SN luminosities
\citep{kelly10,sullivan10}. Several novel methodologies were developed for the analysis of greater and higher
quality datasets, and improve the usefulness of SNe~Ia as distance indicators, as will be summarized below.

\smallskip
{\it Datasets and Methodologies}.  The selection of the methodologies employed for this study is driven by the
{\it a priori} decision to not alter the original analysis of the distant SNe performed by their
authors and consistently apply such formalisms to the nearby SNe. Given this constraint we are able to employ
six prescriptions for the standardization of the SN luminosities. For four of them we calculate ourselves the standardized peak magnitudes for the nearby SNe, and for two methodologies such magnitudes are available in the literature:

\begin{itemize}
\item the \citet[H96]{hamuy96b} technique which used  $BVI$ photometry for a subsample of 26 Calán-Tololo distant SNe
\item the \citet[P99]{phillips99} approach which employed $BVI$ light curves for a subsample of 40 Calán-Tololo+CfA distant SNe
\item the \citet[F10]{folatelli10} implementation based on $JH$ photometry from 31 CSP distant SNe
\item the \citet[K12]{kattner12} model which is based on $JH$ data for the 24 best-observed CSP distant SNe
\item the \citet[F19]{freedman19} method based on the $BiJH$ photometry from 99 CSP distant SNe
\item the \citet[R19]{riess19} method based on the $u^\prime$$g^\prime$$r^\prime$$i^\prime$$UBVRI$ Supercal dataset, a combination of 217 CSP, LOSS, and CfA SNe. 
\end{itemize}

\smallskip
Table \ref{methods} summarizes the six prescriptions employed for this work. In the case of the first four methods (\citetalias{hamuy96b}, \citetalias{phillips99}, \citetalias{folatelli10}, \citetalias{kattner12}), we remeasure the light curve parameters for each of the nearby SNe, such as peak magnitude, colour, and decline rate {\it directly} from the data, that is, without attempting to apply a light curve fitter. These parameters are obtained from a  simple Legendre polynomial fit performed around maximum light and the scatter around the fit yields the peak magnitude error (with an adopted minimum of 0.02 mag). The magnitude decline, \dm, is computed by interpolating the $B$ magnitude directly from the data at an epoch of 15 days past maximum, and 
subtracting the $B$ peak magnitude. A minimum uncertainty of 0.04 mag is adopted for \dm. In this manner we maintain a uniform method of calibration. In these first four papers, a calibration recipe is provided to correct the peak magnitude to a standard candle (absolute magnitude) value. We use this calibration as given in these papers but apply it to our directly measured light curve parameters. In all cases we apply Galactic reddening corrections from \citet[SF11]{schlafly11}, although \citetalias{hamuy96b} used \citet[BH82]{burstein82}, while \citetalias{phillips99} employed \citet[SFD98]{schlegel98}. To ensure internal consistency, we compute the differences among \citetalias{burstein82}, \citetalias{schlegel98} and \citetalias{schlafly11} for each of the samples of distant SNe and apply the corresponding corrections. All of the technical details employed for computing the standardized absolute magnitudes can be found in the Appendix, for each of these four methods.

\begin{table}
\centering
\caption{Prescriptions for standardizing SN magnitudes \label{methods}}
\begin{tabular}{lccc}
\hline
Method &
Reference &
Bandpasses &
Number of distant SNe \\
\hline
\citetalias{hamuy96b}    & \citet{hamuy96b}      & $BVI$ &  26 \\
\citetalias{phillips99}  & \citet{phillips99}    & $BVI$ &  40 \\
\citetalias{folatelli10} & \citet{folatelli10}   & $JH$  &  31 \\
\citetalias{kattner12}   & \citet{kattner12}     & $JH$  &  24 \\
\citetalias{freedman19}  & \citet{freedman19}    & $BiJH$ &  99 \\
\citetalias{riess19}     & \citet{riess19}       & $u^\prime$$g^\prime$$r^\prime$$i^\prime$$UBVRI$ & 217 \\
\hline
\end{tabular}
\end{table}

\smallskip
Table \ref{H96a_data} presents the resulting light curve parameters for the nearby SNe with TRGB distances, for which we are
able to apply the \citetalias{hamuy96b} method.  For each SN we present their $BVI$ absolute magnitudes standardized to an equivalent decline rate of \dm=1.1. The uncertainties quoted for the individual absolute magnitudes are, by choice, the quadrature sum of the uncertainties in the measured parameters and the standardization coefficients, without attempting to estimate systematic errors. Table \ref{H96b_data} presents the same information for the nearby SNe with Cepheid distances, for which we are able to apply the \citetalias{hamuy96b} method. Likewise, the pair of Tables \ref{P99a_data} and \ref{P99b_data} present the results for the $BVI$ filters using the \citetalias{phillips99} method. In Tables \ref{F10a_data} and \ref{F10b_data} we present the results for the $JH$ filters applying the \citetalias{folatelli10} technique. Similarly, Tables \ref{K12a_data} and \ref{K12b_data} summarize the same but using the \citetalias{kattner12} method. At the bottom of these tables we provide, for each filter, the weighted mean absolute magnitude for the whole sample of nearby SNe that we are able to employ in each case, the weighted standard deviation, the standard error of the mean, the error of the weighted mean, and the number of SNe employed\footnote{Not surprisingly, the dispersion and mean error decreased over time from \citetalias{hamuy96b} to  \citetalias{kattner12} as a result improvements in the photometry.}. In the following analysis we also use mean absolute magnitudes for different subsamples of the nearby SNe listed in such tables.

\smallskip

For the remaining two cases, namely, \citetalias{freedman19} and \citetalias{riess19}, the standardized peak magnitudes of the nearby SNe were derived
with a light curve fitter by their own authors. They qualify for this study because both the nearby and distant SN corrected peak
magnitudes were analyzed in a consistent manner and the data are publicly available. In the Appendix we summarize each of these
two methodologies and the relevant parameters drawn from each of them.

\smallskip
Since we apply each of these six prescriptions to one or more bandpasses, we are able to study a total of 12 methodology/bandpass combinations, namely, H96($B$), H96($V$), H96($I$), P99($B$), P99($V$), P99($I$), F10($J$), F10($H$), K12($J$), K12($H$), F19($B$), and R19($B$). In the case of \citetalias{hamuy96b} we derive two sets of solutions, as explained in the Appendix, raising to 15 the number of cases studied. For each of these possible combinations we compute absolute magnitudes and \Ho\ values. We perform this analysis independently for the TRGB and Cepheid calibrations, that is, we obtained a total of 30 sets of absolute magnitudes and \Ho\ values. For each of these methodology/bandpass combinations we attempt to include in the analysis as many of the 18 and 19 nearby SNe with with TRGB and Cepheid distances, respectively. But in some cases, we have to exclude objects that lack the indispensable data required for standardizing their magnitudes in an identical manner as their corresponding distant counterparts. Given that each methodology draws a different sample of nearby SNe, the resulting absolute magnitudes and \Ho\ values are subject to different systematic uncertainties. Hence, care has to be exercised when comparing these techniques, as explained below.

\subsection{Rung 2} \label{rung2}

\begin{figure*}
    \centering
    \includegraphics[width=10cm]{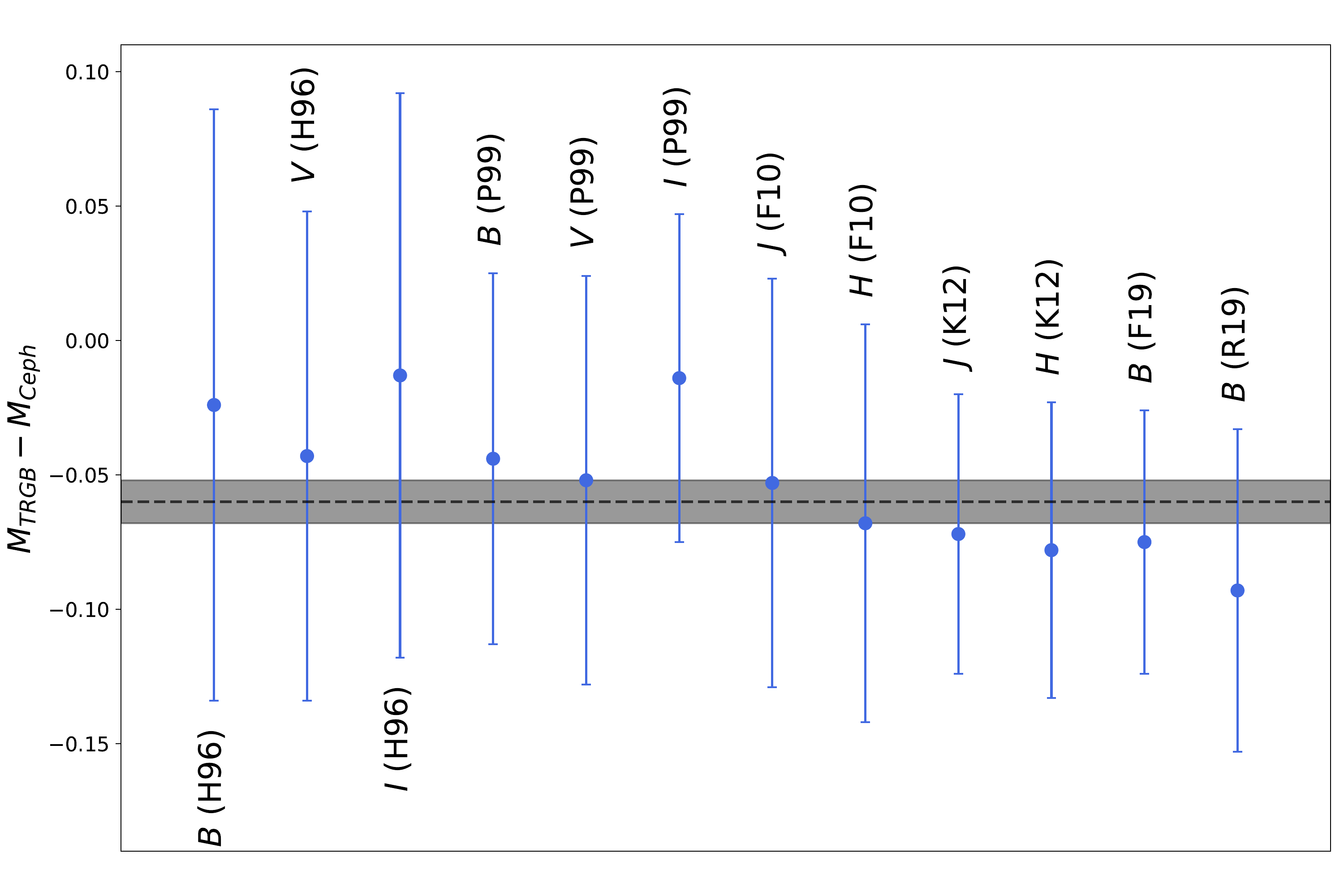}
    \caption{Absolute magnitude differences between the TRGB and Cepheid methods using 12 methodology/bandpass combinations. Note that in all cases the difference is negative, that is, the TRGB method results in brighter magnitudes. The average difference $M_{\mathrm{TRGB}}$-$M_{\mathrm{Ceph}}$=-0.060$\pm$0.008 ($\sigma/\sqrt{n}$) is shown as the grey band.} 
    \label{fig.absmag}
\end{figure*}

In this section we investigate the net effect on absolute magnitudes of SN~Ia, either by adopting the TRGB or Cepheid distances.
We approach this test separately for each of the 12 standardization methodology/bandpass combinations. With the purpose of separating from this test the systematics arising from drawing different nearby SNe from their parent population, for each methodology/bandpass combination we identify the same sample of nearby SNe having both Cepheid and TRGB distances. We obtain between five and ten SNe in common for each methodology/bandpass combination. We then calculate the absolute magnitude difference between the Cepheid and TRGB calibrations. 
Figure \ref{fig.absmag} presents the difference in absolute magnitude for each of the 12 methodology/bandpass combinations. In all cases the absolute magnitudes are systematically brighter when using the TRGB distances. On average the 12 combinations yield $M_{\mathrm{TRGB}}$-$M_{\mathrm{Ceph}}$=-0.060$\pm$0.008 ($\sigma/\sqrt{n}$). Since we use TRGB and Cepheid distance
moduli anchored to the same R1 calibration, this comparison provides a direct estimate of the systematic difference
between the TRGB and Cepheid distance moduli. This difference can be compared to the systematic uncertainties in the TRGB and Cepheid methods, of 0.033 mag \citep{freedman19} and 0.030 mag \citep{riess19}, respectively (excluding the systematic uncertainty in the adopted LMC distance modulus). Adding in quadrature both terms we would expect a 0.044 mag difference in $M_{\mathrm{TRGB}}$-$M_{\mathrm{Ceph}}$, somewhat smaller than our derived value of $M_{\mathrm{TRGB}}$-$M_{\mathrm{Ceph}}$=-0.060$\pm$0.008, thus suggesting that the systematic uncertainties calculated by \citet{freedman19} and/or \citet{riess19} might be somewhat underestimated.

\subsection{Rung 3} \label{rung3}

Here we investigate the robustness of R3 of the CDL using the various surveys and techniques employed to deliver standardized SN luminosities for SNe Ia in the Hubble flow. As mentioned above, the scope of this work focuses on six prescriptions that make use of several distant samples of SNe~Ia such as the Calán-Tololo, CfA, LOSS, and CSP.

\smallskip
The common approach in those analysis has been to establish the redshift-magnitude relationship, a.k.a.
the Hubble diagram. Here the meaning of magnitude is a standardized peak brightness of a SN. Within the context of
the Friedman-Lema\^itre cosmological model, the redshift-magnitude relationship takes the form,

\begin{equation}
        m_{\mathrm{\sc{MAX},corr}} = 5~log~x + \mathrm{ZP},
\label{eq:HD}
\end{equation}
\smallskip

\noindent where $x=d_LH_0$, $d_L$ is the luminosity distance of each SN, \Ho\ is the Hubble-Lema\^itre constant, and ZP is an
empirically determined zero-point provided by the data. In the low redshift ($z$ $<$ 0.1) regime, the redshift-magnitude
relationship can be approximated by a simple kinematical model including an acceleration term,

\begin{equation}
        m_{\mathrm{\sc MAX,corr}} \approx 5~log~\left(cz(1+\frac{1-q_0}{2}z)\right) + \mathrm{ZP},
\end{equation}
\smallskip

\noindent where $q_0$ is the deceleration parameter. Within the aforementioned cosmological framework, ZP relates two physical quantities, \Ho\ and $M_{\mathrm{{\sc MAX},corr}}$, in this simple way:

\begin{equation}
        \mathrm{ZP} = M_{\mathrm{{\sc MAX},corr}} - 5~log~H_0 + 25,
\label{eq:ZP}
\end{equation}
\smallskip

\noindent where $M_{\mathrm{{\sc MAX},corr}}$ is the standardized absolute peak magnitude of SNe~Ia. Hence, the empirically derived ZP of the Hubble diagram interacts directly with the Hubble-Lema\^itre constant, and the peak absolute magnitude is the contact point between ZP and \Ho.

\begin{table}
\centering
\caption{Hubble Diagram Zero Points\label{ZP_table}}
\begin{tabular}{lcc}
\hline
Method &
Published Zero Point (ZP$^\prime$) &
Zero Point (ZP)$^{a}$ \\
\hline
H96(B)            &  -3.318$\pm$0.035     & -3.384$\pm$0.035 $^{b}$  \\
H96(V)            &  -3.329$\pm$0.031     & -3.379$\pm$0.031 $^{b}$  \\
H96(I)            &  -3.057$\pm$0.035     & -3.087$\pm$0.035 $^{b}$ \\
P99(B)            &  28.671$\pm$0.043     & -3.671$\pm$0.043 $^{c}$  \\
P99(V)            &  28.615$\pm$0.043     & -3.615$\pm$0.043 $^{c}$  \\
P99(I)            &  28.236$\pm$0.037     & -3.236$\pm$0.037 $^{c}$  \\
F10(J)            & -18.44$\pm$0.01       & -2.727$\pm$0.01  $^{d}$  \\
F10(H)            & -18.38$\pm$0.02       & -2.667$\pm$0.02  $^{d}$  \\
K12(J)            & -18.552$\pm$0.002     & -2.839$\pm$0.002 $^{d}$  \\
K12(H)            & -18.390$\pm$0.003     & -2.677$\pm$0.003 $^{d}$  \\
F19(B)            & -19.162$\pm$0.010     & -3.449$\pm$0.010 $^{d}$  \\
R16(B)            & -0.71273$\pm$0.00176  & -3.564$\pm$0.009 $^{e}$  \\
\hline
\end{tabular}
\begin{tablenotes}
\item $^{a}$ $m_{\mathrm{MAX,corr}} = 5~log x + \mathrm{ZP}$.
\item $^{b}$ corrected for new Galactic Extinction Calibration, see  section \ref{H96_H0}.
\item $^{c}$ $\mathrm{ZP} = 25 - \mathrm{ZP}^\prime$.
\item $^{d}$ $\mathrm{ZP} = \mathrm{ZP}^\prime + 25 - 5 \times log(72)$.
\item $^{e}$ $\mathrm{ZP} = 5\times \mathrm{ZP}^\prime$.
\end{tablenotes}
\end{table}

\smallskip
Each one of the six prescriptions selected for this work have different definitions for the zero points.
Table \ref{ZP_table} summarizes the original zero points published by their authors (ZP$^\prime$), each of
which is unique for each methodology and band-pass. By choice, we do not to modify them but, we convert
all of them to the definition given in equation \ref{eq:HD} in order to facilitate their comparison. Combining these zero points with the corresponding absolute magnitudes, we proceed to compute \Ho\ values using equation \ref{eq:ZP}. 

\begin{table}
\centering
\caption{Values of \Ho\ in \kmsmpc \label{H0_table}}
\begin{tabular}{lcccccc}
\hline
\multicolumn{1}{l}{Method} &
\multicolumn{1}{c}{H$_0$(B)} &
\multicolumn{1}{c}{H$_0$(V)} &
\multicolumn{1}{c}{H$_0$(I)/H$_0$(i)} &
\multicolumn{1}{c}{H$_0$(J)} &
\multicolumn{1}{c}{H$_0$(H)} &
\multicolumn{1}{c}{TRGB/CEPH} \\
\multicolumn{1}{l}{} &
\multicolumn{1}{c}{$\pm$stat} &
\multicolumn{1}{c}{$\pm$stat} &
\multicolumn{1}{c}{$\pm$stat} &
\multicolumn{1}{c}{$\pm$stat} &
\multicolumn{1}{c}{$\pm$stat} &
\multicolumn{1}{c}{} \\
\hline
H96 (no colour correction)    & 72.9 $\pm$2.7 & 71.3 $\pm$2.2 & 69.8 $\pm$2.6 & --            & --            &  TRGB \\
H96 (with colour correction)  & 66.6 $\pm$2.5 & 66.6 $\pm$2.1 & 67.1 $\pm$2.5 & --            & --            &  TRGB \\
P99                           & 70.2 $\pm$2.0 & 70.1 $\pm$1.9 & 68.7 $\pm$1.8 & --            & --            &  TRGB \\
F10                           & --            & --            & --            & 66.5 $\pm$1.6 & 69.4 $\pm$2.1 &  TRGB \\
K12                           & --            & --            & --            & 69.2 $\pm$1.2 & 70.3 $\pm$1.6 &  TRGB \\
F19 (Tripp method)            & 70.0 $\pm$1.0 & --            & --            & --            & --            &  TRGB \\
R19                           & 70.4 $\pm$1.2 & --            & --            & --            & --            &  TRGB \\
                              &               &               &               &               &               &       \\
H96 (no colour correction)    & 77.0 $\pm$2.6 & 76.3 $\pm$2.1 & 72.5 $\pm$2.7 & --            & --            &  CEPH \\
H96 (with colour correction)  & 72.4 $\pm$2.5 & 72.8 $\pm$2.0 & 70.6 $\pm$2.6 & --            & --            &  CEPH \\
P99                           & 75.0 $\pm$2.0 & 75.4 $\pm$1.9 & 72.4 $\pm$1.8 & --            & --            &  CEPH \\
F10                           & --            & --            & --            & 69.1 $\pm$1.3 & 74.4 $\pm$1.6 &  CEPH \\
K12                           & --            & --            & --            & 72.7 $\pm$1.0 & 75.2 $\pm$1.2 &  CEPH \\
F19 (Tripp method)            & 72.4 $\pm$1.1 & --            & --            & --            & --            &  CEPH \\
R19                           & 73.8 $\pm$1.2 & --            & --            & --            & --            &  CEPH \\
\hline
\end{tabular}
\end{table}

\begin{figure*}
    \centering
    \includegraphics[width=8cm]{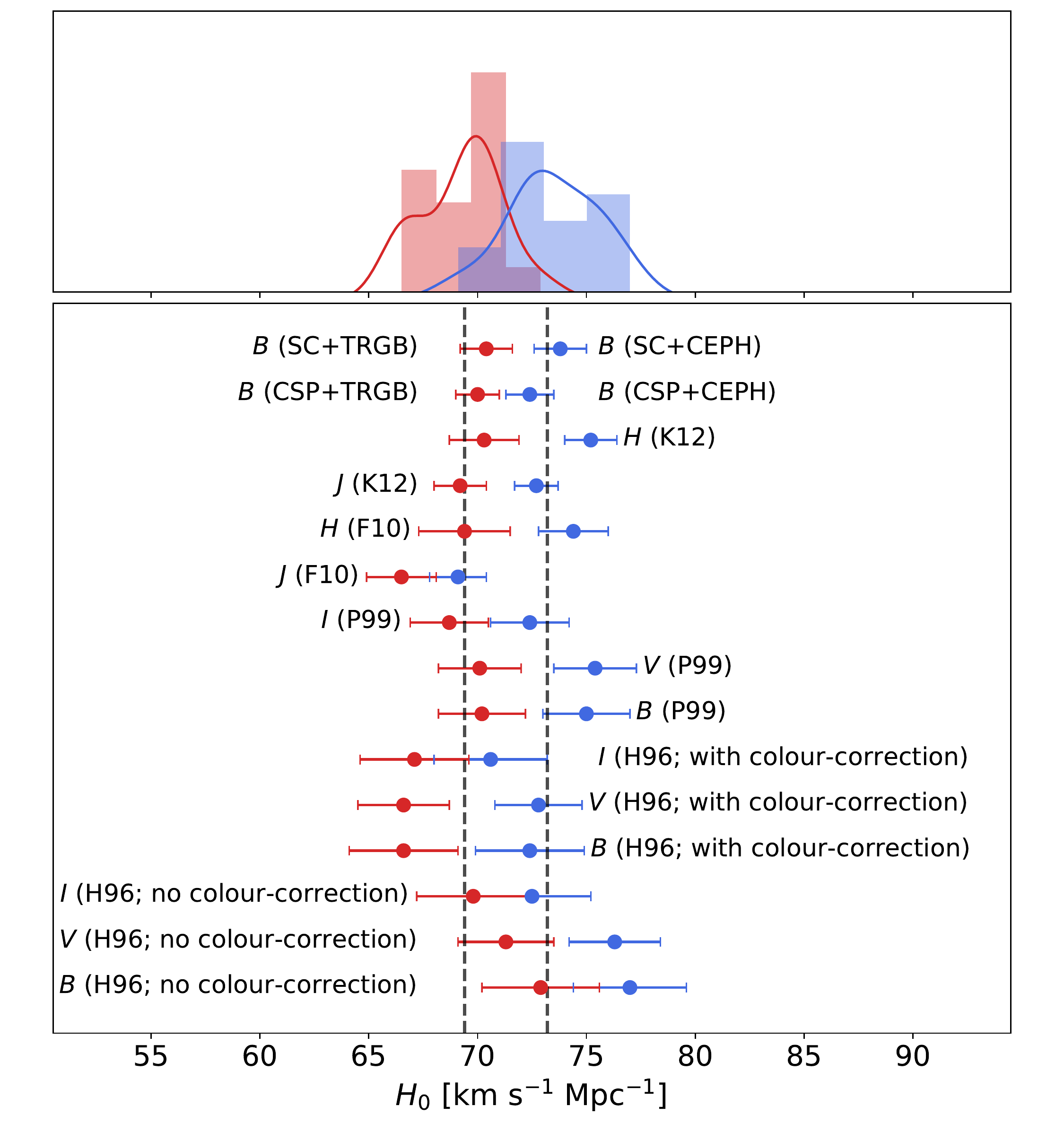}
    \caption{Red points correspond to \Ho\ values derived from 15 different methodology/bandpass combinations calibrated using TRGB distances, with a weighted average and standard deviation of 69.4$\pm$1.9 \kmsmpc. Blue points correspond to \Ho\ values derived from 15 different methodology/bandpass combinations calibrated with Cepheid distances, having a weighted average and standard deviation of 73.2$\pm$2.1 \kmsmpc. Black vertical dashed lines represent weighted average values, and their uncertainties correspond to one $\sigma$.}
    \label{fig.H0.all}
\end{figure*}

\smallskip
In Table \ref{H0_table} and Figure \ref{fig.H0.all} we present the 30 \Ho\ values derived from the 12 methodology/bandpass combinations, using both the TRGB and the Cepheid calibrations, as described in Appendix \ref{appendix}.
Given the systematic differences found for R2 in section \ref{rung2}, we present the TRGB and Cepheid with different colors. As anticipated, there is a clear offset between both distributions. From the TRGB calibration we obtain a weighted average of 69.4$\pm$1.9 ($\sigma$) \kmsmpc. Looking in more detail to the distribution, we note that the most discrepant value is F10(J) with 66.5$\pm$1.6, which lies 1.8 $\sigma$ from the mean. Interestingly, the recalibration by \citetalias{kattner12} gives a value of 69.2$\pm$1.2, and lies comfortably close to the mean value, which suggests that the F10(J) value may be subject to a significant systematic uncertainty. Although the \citetalias{hamuy96b} values derived with no colour corrections are formally consistent with the average, they tend to lie on the high side of the distribution, with a systematic decrease from the $B$, $V$, and $I$ bands. This trend disappears when using the colour-corrected values \footnote {This improvement is expected due to the fact that the original \citetalias{hamuy96b} analysis did not apply host-galaxy reddening corrections to individual SNe but only the removal of suspicious SNe having near-maximum colour $(B_{\mathrm{MAX}}-V_{\mathrm{MAX}})$ $>$ 0.2, that is, those most likely affected by host reddening. This simple colour cutoff leaves little room for significant extinction on the parent galaxies but may introduce a luminosity bias due to unaccounted differential host-galaxy extinction between the distant and the nearby samples. The application of a global colour correction between both samples is a statistical approach that helps to reduce such bias, as clearly shown in Figure \ref{fig.H0.all}.}. The Cepheid calibration yields a weighted average of 73.2$\pm$2.1 ($\sigma$) \kmsmpc. As in the TRGB distribution, we note again that the most discrepant value is F10(J) with 69.1$\pm$1.3, which lies 3.2 $\sigma$ from the mean, but the $J$ band recalibration by \citetalias{kattner12} provides a value of 72.7$\pm$1.0, solving this issue. We note again that the \citetalias{hamuy96b} methodology behaves better when using colour-corrected values.

\smallskip

\begin{figure*}
    \centering
    \includegraphics[width=8cm]{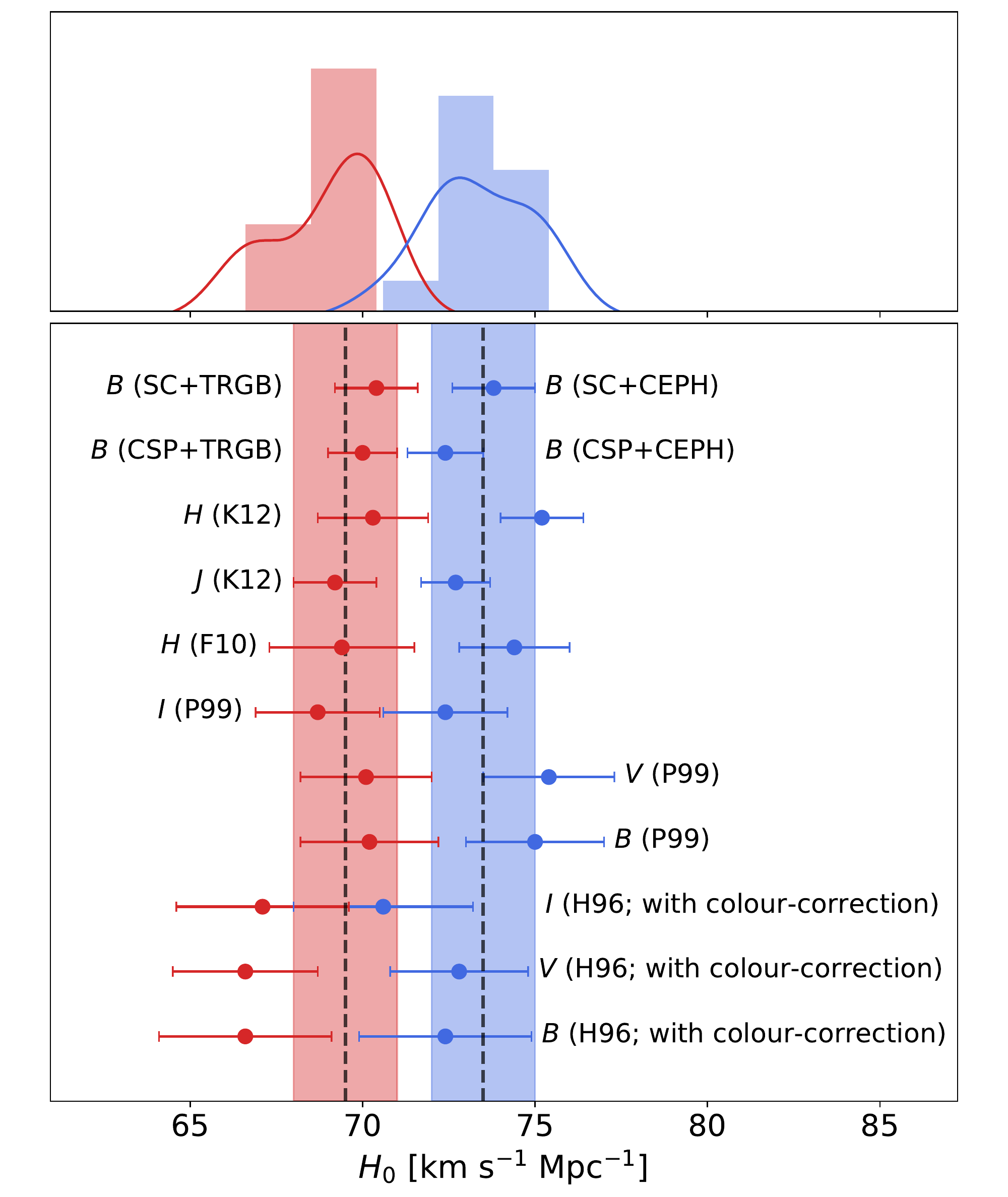}
    \caption{Same as Figure \ref{fig.H0.all} but excluding \citetalias{hamuy96b} values with no colour correction and the F10(J) values. This subset of 22 values yields a weighted average of 69.5$\pm$1.5 \kmsmpc\ for the TRGB calibration and 73.5$\pm$1.5 \kmsmpc\ for the Cepheid calibration. The black vertical dashed lines represent weighted averages, and the blue and red regions correspond to 1 $\sigma$ uncertainties for TGRB and Cepheid calibrations, respectively.}
    \label{fig.H0.all-4}
\end{figure*}

Based on the previous analysis, we show in Figure \ref{fig.H0.all-4} our results but we eliminate the suspicious values, that is, the six \citetalias{hamuy96b} values derived with no color correction and the two F10(J) values. From this subset of 11 methodologies/bandpass we obtain similar averages but with smaller standard deviations, namely,
69.5$\pm$1.5 ($\sigma$) \kmsmpc\ for the TRGB calibration and 73.5$\pm$1.5 ($\sigma$) \kmsmpc\ for the Cepheid calibration.
Adding in quadrature the systematic uncertainty in the TRGB method of 0.033 mag \citep{freedman19} and in the Cepheid technique of 0.030 mag \citep{riess19} (excluding the systematic uncertainty in the adopted LMC distance modulus), the systematic offset between the TRGB and Cepheid calibrations can be clearly seen, with a significance of 1.6 $\sigma$.

\smallskip
We can now go a step further and attempt to measure the error in the mean for each of the two distributions. However, given that several of the 11 methodologies/bandpass combinations considered above do not use entirely independent data, the resulting \Ho\ values are not fully independent from each other, thus implying that the error on the mean cannot be blindly computed from the 11 values. To get around this issue, we select the three most independent possible methodologies/bandpasses: P99(V), K12(J), and R19(B). The first two datasets are fully independent as they do not have any SN in common in the Hubble flow used to determine the ZP. The last two datasets are not fully independent, but only 11\% of the Supercal sample employed by \citetalias{riess19} overlap with the CSP sample used by \citetalias{kattner12}. We reproduce such values in Table \ref{Ho}, and we present the weighted mean and the error in the mean. For the TRGB calibration we obtain \Ho=69.9$\pm$0.8, while for the Cepheid method  we derive \Ho=73.5$\pm$0.7 \kmsmpc.
Adding in quadrature the systematic uncertainty in the TRGB method of 0.033 mag \citep{freedman19} and in the Cepheid technique of 0.030 mag \citep{riess19} (excluding the systematic uncertainty in the adopted LMC distance modulus), this exercise reveals a significant 2.0 $\sigma$ systematic difference in the calibration of R2.
However, if R1 and R2 are held fixed, the different formalisms developed for standardizing the SN peak magnitudes yield consistent results. This study demonstrates that SNe\,Ia have provided a remarkably robust calibration of R3 for over 25 years!

\begin{table}
\centering
\caption{Selected Values of the Hubble-Lema\^itre constant \label{Ho}}
\begin{tabular}{lcc}
\hline
Method & 
\Ho &
TRGB/CEPH \\
&
\kmsmpc &
\\
\hline
P99(V)                             & 70.1$\pm$1.9 & TRGB  \\
K12(J)                             & 69.2$\pm$1.2 & TRGB  \\
R19(B)                             & 70.4$\pm$1.2 & TRGB  \\
\hline
Weighted Mean                      & 69.9$\pm$0.8 & TRGB  \\
\hline
\hline
P99(V)                             & 75.4$\pm$1.9 & CEPH  \\
K12(J)                             & 72.7$\pm$1.0 & CEPH  \\
R19(B)                             & 73.8$\pm$1.2 & CEPH  \\
\hline
Weighted Mean                      & 73.5$\pm$0.7 & CEPH  \\
\hline

\end{tabular}
\end{table}

\begin{figure*}
    \centering
    \includegraphics[width=10cm]{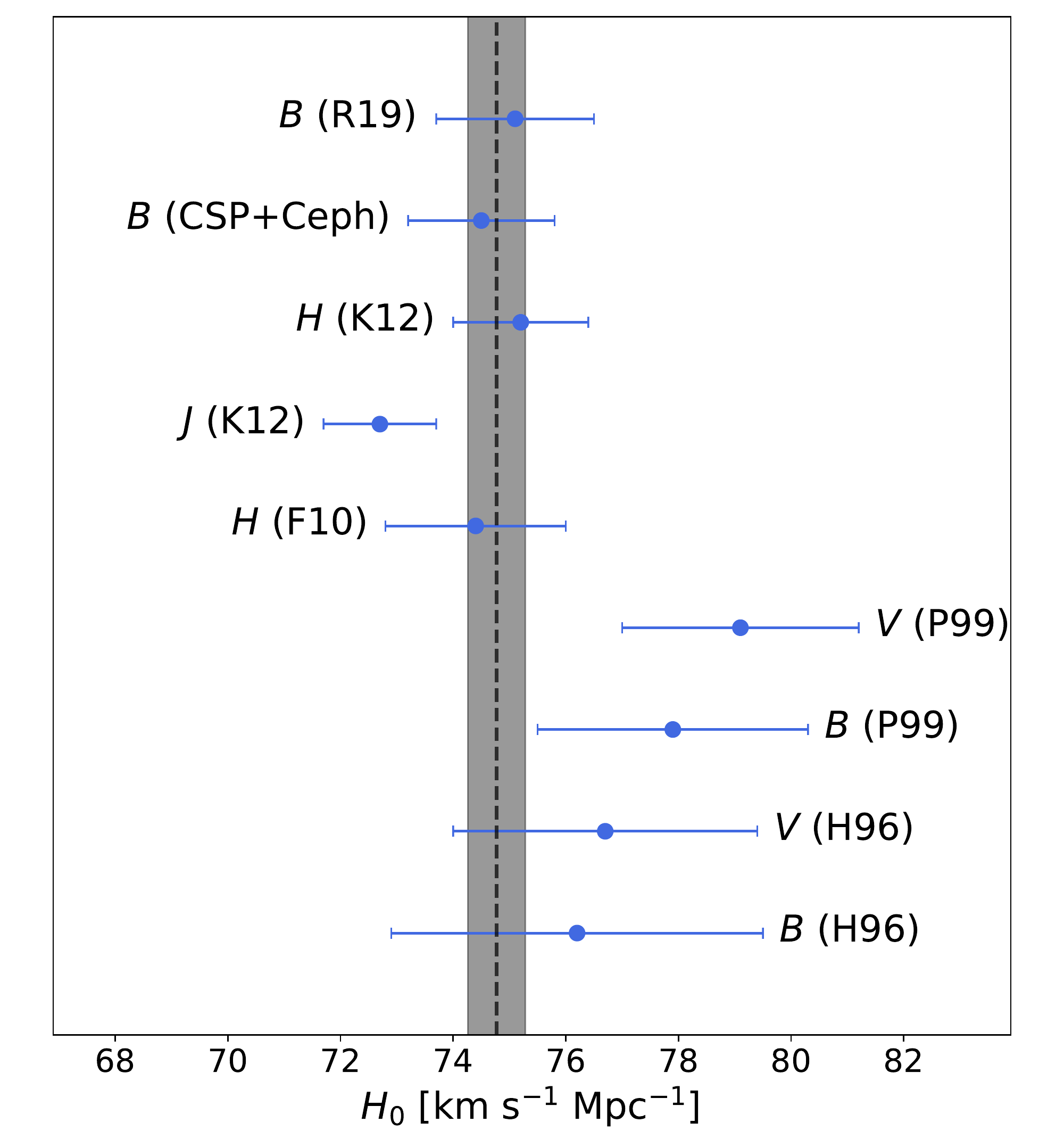}
    \caption{\Ho\ values derived from nine methodology/bandpass combinations, for each of which we used the same common sample
      of nine nearby SNe calibrated with Cepheid distances. The average value of 74.8$\pm$2.0 \kmsmpc\  is shown as the vertical dashed line and the 1 $\sigma$ uncertainty is represented by the grey region.}
    \label{fig.H0.CEPH.n9}
\end{figure*}

\smallskip
We turn now to the challenge of estimating the systematic error in \Ho\ based on SNe~Ia, taking advantage of the large number of methodology/bandpass combinations presented in this study. To address this issue we compare \Ho\ values derived from the {\it same subset of nearby objects}. This approach allows us to isolate the systematics of these formulations from those introduced by the sample of nearby SNe that each methodology draws from the parent population of nearby SNe. We purposely exclude from this study the F10(J) value as well as those obtained using \citetalias{hamuy96b} and no colour correction for the reasons mentioned above. With such constraints we are able to carry out this test using nine SNe in common to nine methodology/bandpass combinations calibrated with Cepheid distances. The \Ho\ values
calculated with these constraints are shown in Figure \ref{fig.H0.CEPH.n9}. The weighted mean \Ho=74.8 has an associated standard deviation of 2.0, which is mainly dominated by the statistical uncertainties of the small sample of nearby SNe (n=9). The $\chi^2_{\nu}$ value of 1.25 indicates that the statistical uncertainties are capable of accounting for most of the dispersion. An small extra uncertainty of 0.2 \kmsmpc\ lowers $\chi^2_{\nu}$ to unity, which can be attributed to systematic uncertainties in these methods. An upper limit to the systematic uncertainties can be estimated from the standard deviation which amounts to 2.0 \kmsmpc, although the majority of it can be attributed to the statistical uncertainties. We repeated the same analysis but using the TRGB distances. In this case the sample of nearby SNe drops to only n=5, the standard deviation is 2.3 \kmsmpc\ and $\chi^2_{\nu}$ is identical to unity.

\section{Conclusions} \label{conclusions}

We assess the robustness of the two highest rungs of the Cosmic Distance Ladder (CDL) for Type Ia supernovae and the corresponding determination of the Hubble-Lema\^itre constant. In this analysis we hold fixed the first rung of the CDL (R1) as the distance modulus to the LMC, 18.48$\pm$0.02, determined to a 1\% precision level using DEB stars \citep{pietrzynski19}. For the second rung (R2) we analyze the two currently most competitive methods, the TRGB and Cepheid luminosity calibration of Type Ia supernovae in nearby galaxies. Finally, for the third rung of the CDL (R3) we analyze various modern digital samples of SNe~Ia in the smooth Hubble flow, such as the Calán-Tololo, CfA, CSP, Supercal datasets, and six prescriptions to standardize their optical and NIR peak luminosities. We apply each of these six prescriptions to one or more bandpasses, leading to a total of 15 determinations of \Ho\ from all possible combinations of bandpasses and methodologies when using the TRGB calibration, and 15 additional determinations for the Cepheid calibration. This metadata analysis allowed us to draw the following conclusions:

\smallskip
$\bullet$ No matter which SN sample, bandpass or methodology is employed for standardizing the SN luminosities, in all cases the \citetalias{freedman19} TRGB calibration yields smaller \Ho\ values than the \citetalias{riess19} Cepheid calibration, a direct consequence of the systematic difference in the distance moduli calibrated from the TRGB and Cepheid methods. From the TRGB calibration we obtain a mean value of \Ho=69.5$\pm$1.5 \kmsmpc\ ($\sigma$), whereas from the Cepheid method we find \Ho=73.5$\pm$1.5 \kmsmpc. Adding in quadrature the systematic uncertainty in the TRGB method of 0.033 mag \citep{freedman19} and in the Cepheid technique of 0.030 mag \citep{riess19} (excluding the systematic uncertainty in the adopted LMC distance modulus), the systematic offset between the TRGB and Cepheid calibrations can be clearly seen, with a significance of 1.6 $\sigma$ (see Fig. \ref{fig.H0.all-4}). 

\smallskip
$\bullet$ Selecting the three most independent possible methodologies/bandpasses (the $V$ band by \citet{phillips99}, the $J$ band by \citet{kattner12}, and the $B$ band by \citet{riess19}), we obtain \Ho=69.9$\pm$0.8 and \Ho=73.5$\pm$0.7 \kmsmpc from the TRGB and Cepheid calibrations, respectively. Adding in quadrature the systematic uncertainty in the TRGB method of 0.033 mag \citep{freedman19} and in the Cepheid technique of 0.030 mag \citep{riess19} (excluding the systematic uncertainty in the adopted LMC distance modulus), this subset reveals a significant 2.0 $\sigma$ systematic difference in the calibration of R2.

\smallskip
$\bullet$ If R1 and R2 are held fixed, the different formalisms developed for standardizing the SN peak magnitudes yield consistent results, with a standard deviation of 1.5 \kmsmpc\, that is, SNe~Ia are able to anchor R3 to a level of 2\% precision. This internal agreement yielded by SNe~Ia, either using the TRGB or Cepheid calibrations, is remarkable as it comprises light curves of increasingly quality, starting with the Calán-Tololo $BVI$ sample, the first digital survey carried out in the early 90s, various releases of the CfA project, and the most modern CSP dataset obtained over recent years with a uniform photometric system over a wide range of optical and NIR bandpasses. This study demonstrates that SNe~Ia have provided a remarkably robust calibration of R3 for over 25 years.

\section*{Acknowledgments}
We thank Chris Burns for his kind and detailed responses to our inquiries about his analysis of the CSP data. We are grateful to Adam Riess and Chuck Bennett for sending us valuable comments on a previous version of this paper, which were incorporated into our study. This research has made use of the NASA/IPAC Extragalactic Database (NED) which is operated by the Jet Propulsion Laboratory, California Institute of Technology, under contract with the National Aeronautics and Space Administration.
MH acknowledges support from the Hagler Institute of Advanced Study at Texas A\&M University. NBS has been supported by the NSF grant AST-1613455 and the Mitchell/Heep/Munnerlyn Chair in Observational Astronomy. Additional support has come from the George P. and Cynthia Woods Mitchell Institute for Fundamental Physics and Astronomy.

\section*{DATA AVAILABILITY}

All data used in this paper are available on request to the corresponding author.

\bibliographystyle{mnras}
\bibliography{references} 

\appendix
\section{}
\label{appendix}

\bigskip


This appendix describes six prescriptions that allow one to calculate standardized absolute peak magnitudes for SNe~Ia. We apply these recipes to the set of nearby SNe that possess either Cepheid or TRGB distances from \citetalias{riess19} and \citetalias{freedman19}, respectively. In the first four cases we employ the published prescription for measuring, in the first place, the standardized apparent peak magnitudes, after which we subtract the corresponding distance modulus. In the last two cases we omit the first step since the standardized apparent magnitudes are available in the literature. In each case we proceed to compute the corresponding values of \Ho\ by combining the absolute magnitudes with the zero point of the Hubble diagram derived, in each case, from SNe~Ia in the Hubble flow.

\subsection{The H96 methodology} \label{H96}

The \citet{hamuy96b} methodology was developed to analyze the sample of 29 distant SNe Ia obtained
in the course of the Calán-Tololo project, which constituted the first sample of SNe in the Hubble flow
observed with modern linear CCD detectors. Maximum light magnitudes in the $BVI$ bands and the decline rate
parameter \dm\ were measured for each SN \citepalias{hamuy96c}. A Hubble diagram was obtained for each band,
after correcting the peak magnitudes for the Galactic reddening provided by \citetalias{burstein82}, K-terms
\citepalias{hamuy93b}, and decline rate \dm. Although the individual SNe were not corrected for
host-galaxy reddening, three outlier objects were removed from the initial sample having
the pseudo-colour $(B_{\mathrm{MAX}}-V_{\mathrm{MAX}})$ $>$ 0.2, that is, those most likely affected by host reddening. This simple
colour cutoff left little room for significant extinction on the parent galaxies. In fact, the weighted average
pseudo-colour of the 26 remaining SNe, 0.007$\pm$0.013 ($\sigma/\sqrt N$), is quite normal for unextinxguished
SNe\,Ia. All of the above led to Hubble diagrams with remarkably low dispersions of 0.17, 0.14, 0.13 mag, in
$B$, $V$, $I$, respectively, thus opening the path to high precision cosmology \citepalias{hamuy96b}. Cepheid
distances measured with HST by Sandage, Saha, and collaborators to the host galaxies of SNe 1937C, 1972E, 1981B, 1990N
\citep{saha94,saha95,saha96,sandage96} allowed the calibration of the Calán-Tololo Hubble diagram and derive
a value of 63$\pm$5 \kmsmpc\ for the Hubble-Lema\^itre constant.

\subsubsection{Absolute Magnitudes} \label{H96_AM}

We use now the \citetalias{hamuy96b} methodology to determine absolute magnitudes for the 18 nearby SNe Ia with TRGB
distances published by \citetalias{freedman19}, in the same manner as done for the distant SNe. A requisite to include
such SNe in the re-analysis of the Calán-Tololo data is that each object must have available photometry in the Landolt
standard photometric system \citep{landolt92}, which means that the reduced magnitudes include a photometric colour term (of course, this color term is not correct for SNe, and an S-correction \citep{stritzinger02} is normally needed, but since we are applying the original \citetalias{hamuy96b} formula, no correction is needed for this purpose). Two of
the nearby SNe, SN 2007on and SN 2007sr, do not fulfill this condition. For the remaining SNe we measure their $BVI$
peak magnitudes directly from the data, using a simple Legendre polynomial, as explained in section \ref{analysis}. Following \citetalias{hamuy96b}, we exclude all SNe with $B_{\mathrm{MAX}}-V_{\mathrm{MAX}}$ $>$ 0.2, namely, SN 1989B and SN 1998bu, which reduced to 14 the number of SNe with TRGB distances. 

Table \ref{H96a_data} summarizes our measurements for such nearby SNe (14 in $B$ and $V$,
and 8 in the $I$ filter), including their TRGB distances, peak magnitudes, decline rate, $E(B-V)$ from the NASA
Extragalactic Database (NED), the source for the photometry, and the SN peak absolute $BVI$ magnitudes corrected for
Galactic reddening and decline rate (to the fiducial value of \dm=1.1). The uncertainty in an individual absolute
magnitude is the result of adding in quadrature the uncertainties in peak magnitude, Galactic extinction,
distance modulus, decline rate, the slope of the peak magnitude-decline rate relation, and an additional term
amounting to 0.05 mag that we attribute to the fact that the SN magnitudes were not corrected for S-terms
\citep{suntzeff00,stritzinger02}. Although the lack of S-correction constitutes a systematic uncertainty for an
individual magnitude, they should tend to behave randomly for the ensemble of data points. 

\smallskip
For each of the $BVI$ bands, we proceed to compute the weighted mean absolute magnitude corrected for \dm\ ($M_{\mathrm{MAX,corr}}$), the weighted standard deviation ($\sigma$), the standard error of the mean ($\sigma_M=\sigma/\sqrt{n}$),
and the error of the weighted mean. Note that the standard deviations for the local SNe are 0.26 mag in $B$,
0.22 in $V$, and 0.20 in $I$, that is, $\sim$0.07 mag greater, in all three bands, than the scatter yielded by the
SNe in the Hubble flow which ranges between 0.17-0.13 mag. Possible explanations for the increase in the scatter
could be due to unaccounted host-galaxy extinction corrections in the nearby sample or uncertainties in the
host galaxies distances. 

\smallskip
In view that the \citetalias{hamuy96b} method applied a simple colour cut to correct the SNe for host-galaxy
extinction, it proves relevant to compare the colours of the nearby SNe with those in the Hubble flow. We analyze first the TRGB sample of 14 nearby SNe. For this dataset the weighted mean $B_{\mathrm{MAX}}-V_{\mathrm{MAX}}$ colour, after correcting for
Galactic extinction \citep{schlafly11}, is 0.037$\pm$0.019 ($\sigma/\sqrt n$). For the distant sample the corresponding color is -0.010$\pm$0.013. It is possible that this difference could be due to unaccounted differential host-galaxy extinction between the distant and the nearby samples. Hence, we decide to compute a mean absolute magnitudes by forcing the nearby sample to have the same bluer colour of the distant sample. This required decreasing the previous $M_{\mathrm{MAX,corr}}$ values by (0.037+0.010)$\times$$A_\lambda$/$E(B-V)$, where $A_B$=4.16, $A_V$=3.14, $A_I$=1.82. Table \ref{H96a_data} 
includes mean absolute magnitudes corrected for colour.

\smallskip
Now we apply the \citetalias{hamuy96b} technique to nearby SNe with Cepheid distances published by the SH0ES program.
Again, to be consistent with \citetalias{hamuy96b} we restrict the sample of nearby SNe to those with $E(B-V)<$0.2
and $BVI$ magnitudes in the Landolt standard system. These two restrictions permit us to apply this method to 17
nearby SNe in $B$ and $V$, and 9 SNe in the $I$ filter. Table \ref{H96b_data} presents the relevant parameters of
the SNe, the distance moduli published by \citetalias{riess16} to which we added a global correction of
-0.029 mag (=log 73.24/74.22) in order to place them in the \citetalias{riess19} Cepheid scale, and the resulting 
mean absolute magnitude corrected for \dm. This set of 17 nearby SNe with Cepheid distances has a mean $B_{\mathrm{MAX}}-V_{\mathrm{MAX}}$ colour, corrected for Galactic extinction \citep{schlafly11}, of 0.022$\pm$0.018 ($\sigma/\sqrt n$), that is, redder than the -0.010$\pm$0.013 color of the distant sample. Table \ref{H96b_data} includes mean absolute magnitudes computed by forcing the nearby sample to have the same color of the distant sample.

\subsubsection{The Hubble-Lema\^itre constant} \label{H96_H0}

Having determined absolute magnitudes, it is straightforward to compute \Ho\ with the formula:

\begin{equation}
        log~H_0 = 0.2~(M_{\mathrm{MAX,corr}} - \mathrm{ZP}' + 25)
\end{equation}

\smallskip
\noindent where, $M_{\mathrm{MAX,corr}}$ is the mean absolute magnitude of the nearby SNe corrected for decline rate
and foreground extinction (given in Tables \ref{H96a_data} and \ref{H96b_data}), and $ZP'$ is the zero-point
of the Hubble diagram.

\smallskip
The zero points of the $BVI$ Hubble diagrams derived by \citetalias{hamuy96b} were -3.318, -3.329, -3.057,
respectively. These values need to be corrected owing to the fact that the Galactic extinction applied by
\citetalias{hamuy96b} to the distant sample \citep{burstein82} differs from the new calibration \citep{schlafly11}
that we use for the nearby SNe. Given that the \citet{burstein82} calibration yielded a mean $E(B-V)$ correction
of 0.031 mag for the ensemble of 26 distant SNe, and the new calibration of \citet{schlafly11} yields a somewhat
greater correction of 0.047 mag, we have to decrease the zero points of the Hubble diagrams to -3.384, -3.379, -3.087
in $B$, $V$, and $I$, respectively.

\smallskip

We analyze first the TRGB sample of nearby SNe. Given the discussion above about the color difference between the nearby and distant samples, we decide to calculate two sets of solutions: one ignoring the colour difference between both samples and one that forces both samples to have the same colour. Without taking into account colour differences we obtain \Ho(B)=72.9, \Ho(V)=71.3, and \Ho(V)=69.8. Correcting for colour differences between the nearby (redder) and distant (bluer) samples, the resulting \Ho\ values are
lower than those derived without correcting for color difference and much more consistent among the three filters: 
66.6, 66.6, and 67.1 for $B$, $V$, and $I$, respectively. If there was significant differential reddening between 
the nearby and distant sample, we should observe a dependence of the \Ho\ value as a function of wavelength, which is not the case. Hence, it is encouraging that the colour correction yields values nearly independent on the band considered. 
The resulting values for \Ho\ are summarized in Table \ref{H0_table}, with and without color correction.

\smallskip
Now we analyze the Cepheid sample of nearby SNe in the same manner as above for the TRGB sample. Without considering colour differences we derive \Ho(B)=77.0, \Ho(V)=76.3, and \Ho(V)=72.5. Forcing both datasets to match the same colour, we obtain \Ho\ values of 72.4, 72.8, and 70.6 for $B$, $V$, and $I$, respectively, which are internally consistent within the statistical uncertainties. The resulting values for \Ho\ are summarized in Table \ref{H0_table}. As can be seen in this table, the values derived using the \citetalias{hamuy96b} methodology are in excellent agreement with those derived from modern and larger datasets such as the CSP or Supercal. The Hubble flow from 1996 was sufficient to derive the modern value of the Hubble-Lema\^itre constant. We only had to wait until a better calibration of the distance to Cepheids and an improved reddening map were made.

\begin{landscape}
\begin{table}
\caption{Parameters for Individual Supernovae for \citetalias{hamuy96b} sample and TRGB distances \label{H96a_data}}
\begin{tabular}{lccccccccccc}
  \hline
  \multicolumn{1}{l}{SN} &
  \multicolumn{1}{c}{Galaxy} &
  \multicolumn{1}{c}{Distance} &
  \multicolumn{1}{c}{$B_{\mathrm{MAX}}$} &
  \multicolumn{1}{c}{$V_{\mathrm{MAX}}$} &
  \multicolumn{1}{c}{$I_{\mathrm{MAX}}$} &
  \multicolumn{1}{c}{$\Delta m_{15}(B)$} &
  \multicolumn{1}{c}{$E(B-V)_{\mathrm{GAL}}$} &
  \multicolumn{1}{c}{$M^B_{\mathrm{MAX,corr}}$} &
  \multicolumn{1}{c}{$M^V_{\mathrm{MAX,corr}}$} &
  \multicolumn{1}{c}{$M^I_{\mathrm{MAX,corr}}$} &
  \multicolumn{1}{c}{Photometry} \\
  \multicolumn{1}{l}{} &
  \multicolumn{1}{c}{Name} &
  \multicolumn{1}{c}{Modulus} &
  \multicolumn{1}{c}{} &
  \multicolumn{1}{c}{} &
  \multicolumn{1}{c}{} &
  \multicolumn{1}{c}{} &
  \multicolumn{1}{c}{$\pm$0.020} &
  \multicolumn{1}{c}{} &
  \multicolumn{1}{c}{} &
  \multicolumn{1}{c}{} &
  \multicolumn{1}{c}{Source} \\
  \multicolumn{1}{l}{} &
  \multicolumn{1}{c}{} &
  \multicolumn{1}{c}{} &
  \multicolumn{1}{c}{} &
  \multicolumn{1}{c}{} &
  \multicolumn{1}{c}{} &
  \multicolumn{1}{c}{} &
  \multicolumn{1}{c}{\citetalias{schlafly11}} &
  \multicolumn{1}{c}{} &
  \multicolumn{1}{c}{} &
  \multicolumn{1}{c}{} &
  \multicolumn{1}{c}{} \\
  \hline
 SN\,1980N  &  N1316   &  31.46(0.04) &   12.49(0.02) & 12.44(0.02)&  12.71(0.10) &  1.28(0.04) &  0.020 & -19.194(0.116)    & -19.210(0.100)   &  -18.890(0.130)  & \citet{hamuy91}     \\
 SN\,1981B  &  N4536   &  30.96(0.05) &   12.03(0.03) & 11.93(0.03)& --           &  1.10(0.07) &  0.017 & -19.001(0.126)    & -19.083(0.111)   &    --            & \citet{buta83}          \\
 SN\,1981D  &  N1316   &  31.46(0.04) &   12.59(0.04) & 12.40(0.04)& --           &  1.44(0.05) &  0.020 & -19.220(0.134)    & -19.363(0.116)   &    --            & \citet{hamuy91}     \\
 SN\,1994D  &  N4526   &  31.00(0.07) &   11.89(0.02) & 11.90(0.02)&  12.06(0.05) &  1.31(0.08) &  0.021 & -19.362(0.142)    & -19.314(0.126)   &  -19.099(0.121)  & \citet{richmond95}  \\
 SN\,1994ae &  N3370   &  32.27(0.05) &   13.19(0.02) & 13.10(0.02)&  13.35(0.03) &  0.86(0.05) &  0.029 & -19.012(0.126)    & -19.091(0.109)   &  -18.835(0.099)  & \citet{riess99}    \\
 SN\,1995al &  N3021   &  32.22(0.05) &   13.38(0.02) & 13.24(0.02)&  13.52(0.05) &  0.83(0.05) &  0.013 & -18.682(0.128)    & -18.830(0.111)   &  -18.568(0.109)  & \citet{riess99}    \\
 SN\,2001el &  N1448   &  31.32(0.06) &   12.84(0.02) & 12.73(0.02)&  12.80(0.04) &  1.15(0.05) &  0.014 & -18.577(0.123)    & -18.669(0.108)   &  -18.574(0.100)  & \citet{krisciunas03}\\
 SN\,2002fk &  N1309   &  32.50(0.07) &   13.30(0.03) & 13.37(0.03)&  13.57(0.03) &  1.02(0.04) &  0.038 & -19.295(0.128)    & -19.193(0.115)   &  -18.953(0.102)  & \citet{cartier14}   \\
 SN\,2006dd &  N1316   &  31.46(0.04) &   12.30(0.05) & 12.27(0.05)&  12.46(0.05) &  1.28(0.05) &  0.020 & -19.384(0.127)    & -19.380(0.112)   &  -19.140(0.099)  & \citet{stritzinger10}  \\
 SN\,2007af &  N5584   &  31.82(0.10) &   13.29(0.02) & 13.25(0.02)& --           &  1.26(0.04) &  0.038 & -18.814(0.147)    & -18.802(0.135)   &    --            & \citet{krisciunas17}     \\
 SN\,2011fe &  M101    &  29.08(0.04) &   10.02(0.04) &  9.96(0.04)&  10.23(0.04) &  1.15(0.04) &  0.008 & -19.132(0.117)    & -19.180(0.102)   &  -18.893(0.087)  & \citet{richmond12}  \\
 SN\,2011iv &  N1404   &  31.42(0.05) &   12.48(0.03) & 12.49(0.03)& --           &  1.77(0.05) &  0.010 & -19.507(0.171)    & -19.435(0.146)   &    --            & \citet{gall18}     \\
 SN\,2012cg &  N4424   &  31.00(0.06) &   12.11(0.03) & 11.99(0.03)& --           &  0.92(0.04) &  0.019 & -18.828(0.126)    & -18.942(0.112)   &    --            & \citet{marion16}    \\
 SN\,2012fr &  N1365   &  31.36(0.05) &   12.02(0.02) & 12.04(0.02)& --           &  0.82(0.04) &  0.020 & -19.204(0.126)    & -19.185(0.109)   &    --            & \citet{contreras18}     \\
\hline
Mean (no color correction)  &  &              &               &            &              &             &        & -19.072           & -19.113          &  -18.866         &                \\
Mean (color correction)     &  &              &               &            &              &             &        & -19.267           & -19.261          &  -18.951         &                \\
$\sigma$                    &  &              &               &            &              &             &        &  0.269            &  0.225           &   0.209          &                \\
$\sigma$/$\sqrt n$          &  &              &               &            &              &             &        &  0.072            &  0.060           &   0.074          &                \\
Error in Mean               &  &              &               &            &              &             &        &  0.035            &  0.030           &   0.037          &                \\
$n$                         &  &              &               &            &              &             &        &  14               &  14              &   8              &                \\
\hline
\end{tabular}
\end{table}
\end{landscape}

\begin{landscape}
\begin{table}
\caption{Parameters for Individual Supernovae for \citetalias{hamuy96b} sample and Cepheid distances \label{H96b_data}}
\begin{tabular}{lccccccccccc}
  \hline
  \multicolumn{1}{l}{SN} &
  \multicolumn{1}{c}{Galaxy} &
  \multicolumn{1}{c}{Distance} &
  \multicolumn{1}{c}{$B_{\mathrm{MAX}}$} &
  \multicolumn{1}{c}{$V_{\mathrm{MAX}}$} &
  \multicolumn{1}{c}{$I_{\mathrm{MAX}}$} &
  \multicolumn{1}{c}{$\Delta m_{15}(B)$} &
  \multicolumn{1}{c}{$E(B-V)_{\mathrm{GAL}}$} &
  \multicolumn{1}{c}{$M^B_{\mathrm{MAX,corr}}$} &
  \multicolumn{1}{c}{$M^V_{\mathrm{MAX,corr}}$} &
  \multicolumn{1}{c}{$M^I_{\mathrm{MAX,corr}}$} &
  \multicolumn{1}{c}{Photometry} \\
  \multicolumn{1}{l}{} &
  \multicolumn{1}{c}{Name} &
  \multicolumn{1}{c}{Modulus} &
  \multicolumn{1}{c}{} &
  \multicolumn{1}{c}{} &
  \multicolumn{1}{c}{} &
  \multicolumn{1}{c}{} &
  \multicolumn{1}{c}{$\pm$0.020} &
  \multicolumn{1}{c}{} &
  \multicolumn{1}{c}{} &
  \multicolumn{1}{c}{} &
  \multicolumn{1}{c}{Source} \\
  \multicolumn{1}{l}{} &
  \multicolumn{1}{c}{} &
  \multicolumn{1}{c}{} &
  \multicolumn{1}{c}{} &
  \multicolumn{1}{c}{} &
  \multicolumn{1}{c}{} &
  \multicolumn{1}{c}{} &
  \multicolumn{1}{c}{\citetalias{schlafly11}} &
  \multicolumn{1}{c}{} &
  \multicolumn{1}{c}{} &
  \multicolumn{1}{c}{} &
  \multicolumn{1}{c}{} \\
  \hline
 SN\,1981B  &  N4536   &  30.877(053) &   12.03(0.03) & 11.93(0.03)& --           &  1.10(0.07) &  0.017 & -18.918(0.127)    & -19.000(0.112)   &    --            & \citet{buta83}           \\
 SN\,1990N  &  N4639   &  31.503(071) &   12.76(0.03) & 12.70(0.02)&  12.94(0.02) &  1.07(0.05) &  0.012 & -18.769(0.130)    & -18.819(0.115)   &  -18.568(0.101)  & \citet{lira98}          \\
 SN\,1994ae &  N3370   &  32.043(049) &   13.19(0.02) & 13.10(0.02)&  13.35(0.03) &  0.86(0.05) &  0.029 & -18.785(0.125)    & -18.864(0.109)   &  -18.608(0.099)  & \citet{riess99}     \\
 SN\,1995al &  N3021   &  32.469(090) &   13.38(0.02) & 13.24(0.02)&  13.52(0.05) &  0.83(0.05) &  0.013 & -18.931(0.148)    & -19.079(0.134)   &  -18.817(0.133)  & \citet{riess99}      \\
 SN\,1998aq &  N3982   &  31.708(069) &   12.35(0.02) & 12.46(0.02)&  12.69(0.02) &  1.09(0.04) &  0.000 & -19.350(0.125)    & -19.241(0.111)   &  -19.012(0.098)  & \citet{riess05}     \\
 SN\,2001el &  N1448   &  31.282(045) &   12.84(0.02) & 12.73(0.02)&  12.80(0.04) &  1.15(0.05) &  0.014 & -18.539(0.116)    & -18.631(0.101)   &  -18.536(0.091)  & \citet{krisciunas03} \\
 SN\,2002fk &  N1309   &  32.494(055) &   13.30(0.03) & 13.37(0.03)&  13.57(0.03) &  1.02(0.04) &  0.038 & -19.289(0.121)    & -19.187(0.106)   &  -18.947(0.092)  & \citet{cartier14}   \\
 SN\,2003du &  U9391   &  32.890(063) &   13.44(0.04) & 13.55(0.04)&  13.84(0.02) &  1.09(0.05) &  0.000 & -19.442(0.129)    & -19.333(0.115)   &  -19.044(0.095)  & \citet{hicken09}   \\
 SN\,2005cf &  N5917   &  32.234(102) &   13.63(0.02) & 13.56(0.02)& --           &  1.03(0.04) &  0.077 & -18.869(0.146)    & -18.866(0.135)   &    --            & \citet{hicken09}     \\
 SN\,2007af &  N5584   &  31.757(046) &   13.29(0.02) & 13.25(0.02)& --           &  1.26(0.04) &  0.038 & -18.751(0.117)    & -18.739(0.102)   &    --            & \citet{krisciunas17}  \\
 SN\,2009ig &  N1015   &  32.468(081) &   13.58(0.04) & 13.46(0.02)& --           &  0.86(0.04) &  0.015 & -18.762(0.143)    & -18.885(0.125)   &    --            & \citet{hicken12} \\
 SN\,2011by &  N3972   &  31.558(070) &   12.94(0.02) & 12.89(0.02)&  12.97(0.02) &  1.07(0.04) &  0.000 & -18.594(0.125)    & -18.647(0.112)   &  -18.571(0.098)  & \citet{stahl19} \\
 SN\,2011fe &  M101    &  29.106(045) &   10.02(0.04) &  9.96(0.04)&  10.23(0.04) &  1.15(0.04) &  0.008 & -19.158(0.119)    & -19.206(0.105)   &  -18.919(0.090)  & \citet{richmond12}  \\
 SN\,2012cg &  N4424   &  31.051(292) &   12.11(0.03) & 11.99(0.03)& --           &  0.92(0.04) &  0.019 & -18.879(0.312)    & -18.933(0.307)   &    --            & \citet{marion16}  \\
 SN\,2012fr &  N1365   &  31.278(057) &   12.02(0.02) & 12.04(0.02)& --           &  0.82(0.04) &  0.020 & -19.122(0.129)    & -19.103(0.113)   &    --            & \citet{contreras18} \\
 SN\,2012ht &  N3447   &  31.879(043) &   13.11(0.02) & 13.10(0.02)& --           &  1.29(0.04) &  0.012 & -18.968(0.118)    & -18.951(0.102)   &    --            & \citet{burns18}  \\
 SN\,2015F  &  N2442   &  31.482(053) &   13.49(0.02) & 13.29(0.02)& --           &  1.43(0.04) &  0.178 & -18.991(0.131)    & -18.984(0.114)   &    --            & \citet{burns18}  \\
\hline
Mean (no color correction)               &  &              &               &            &              &             &        & -18.952           & -18.967          &  -18.784         &                \\
Mean (color correction)                  &  &              &               &            &              &             &        & -19.085           & -19.067          &  -18.842         &                \\
$\sigma$                                 &  &              &               &            &              &             &        &  0.267            &  0.214           &   0.214          &                \\
$\sigma$/$\sqrt n$                       &  &              &               &            &              &             &        &  0.065            &  0.052           &   0.071          &                \\
Error in Mean                            &  &              &               &            &              &             &        &  0.032            &  0.028           &   0.033          &                \\
$n$                                      &  &              &               &            &              &             &        &  17               &  17              &   9              &                \\
\hline
\end{tabular}
\end{table}
\end{landscape}

\subsection{The P99 methodology} \label{P99}

The \citet{phillips99} methodology improved the previous work by \citetalias{hamuy96b} by determining host-galaxy
reddening to individual SNe through three novel independent methods: one based on the fact that the $B-V$ colour
30-90 days past $V$ maximum evolve in a similar manner for most SNe Ia use \citep[a.k.a the ``Lira Law'',][]{lira96}
a second one using a calibration of the $B_{\mathrm{MAX}}-V_{\mathrm{MAX}}$ colour with \dm, and a third that calibrates the
$V_{\mathrm{MAX}}-I_{\mathrm{MAX}}$ colour with \dm. These techniques were tested using 62 SNe: 29 from the Calán-Tololo project,
20 objects from the CfA work \citep{riess99}, and 13 well-observed nearby SNe, whose peak magnitudes had been
previously corrected for Galactic extinction using the calibration of \citet{schlegel98}, and for K terms
\citep{hamuy93b}.

\smallskip
When applied to a sample of 17 ``low host-galaxy reddening'' SNe with decline rates of 0.85 \textless \dm \textless 1.7,
a well-behaved peak magnitude-decline rate relation emerged, which was modeled with a quadratic function of
the form \dm\ = a [\dm\ -1.1] + b [\dm\ -1.1]$^2$ with dispersions of 0.11, 0.09, and 0.13 mag in $BVI$, respectively,
clearly lower than the ones obtained by \citetalias{hamuy96b} in the $BV$ bands.

\smallskip
After applying these corrections due to host-galaxy reddening to the 40 SNe in the Hubble flow ($z$ \textgreater 0.01),
\citetalias{phillips99} obtained Hubble diagrams in the $BVI$ bands, with dispersions of $\sim$0.14 mag. The resulting $BVI$
Hubble diagrams were combined with the six SN peak magnitudes calibrated with Cepheid distances \citep{saha99,suntzeff99},
which led to a value of \Ho=63.3$\pm$2.2$\pm$3.5 \kmsmpc. 

\subsubsection{Absolute Magnitudes} \label{P99_AM}

Now we apply the \citetalias{phillips99} technique to nearby SNe with TRGB distances. To be consistent with
\citetalias{phillips99} we restrict the sample of nearby SNe to those meeting the following two requirements:
(1) having $BVI$ photometry in the Landolt standard photometric system, and (2) lying in the range 0.85
\textless \dm \textless 1.7. This restriction permits us to apply this method to 15 nearby SNe in $B$ and $V$,
and 10 SNe in the $I$ filter. 

\smallskip
We follow the same procedure described in \citetalias{phillips99}, that is, we measure peak magnitudes, decline rates,
and host-galaxy reddening directly from the light curves (in the same manner described above in \ref{H96_AM}), which are
summarized in Table \ref{P99a_data}. 
The mean magnitudes for the ensemble of SNe (shown at the bottom of Table \ref{P99a_data}) are characterized by dispersions between 0.14-17 mag, that is, 0.04 mag greater than those yielded by the distant samples, possibly due to uncertainties in the TRGB distances.

\smallskip
Now we apply the \citetalias{phillips99} technique to nearby SNe with Cepheid distances, restricting the sample to those SNe
with $BVI$ magnitudes in the Landolt standard system lying in the range 0.85 \textless \dm \textless 1.7. This restriction
permits us to apply this method to 18 nearby SNe in $B$ and $V$, and 10 SNe in the $I$ filter. Table \ref{P99b_data} presents
the relevant parameters of the SNe along with the distance moduli published by \citetalias{riess16} to which we add a global
correction of -0.029 mag (= 5 log 73.24/74.22) in order to place them in the \citetalias{riess19} Cepheid scale.

\subsubsection{The Hubble-Lema\^itre constant} \label{P99_H0}

For the \citetalias{phillips99} implementation the value of \Ho\ can be obtained with the formula:

\begin{equation}
        log~H_0 = 0.2~(M_{\mathrm{MAX,corr}} + \mathrm{ZP}')
\end{equation}

\smallskip
\noindent where, $M_{\mathrm{MAX,corr}}$ is the mean absolute magnitude of the nearby SNe corrected for decline rate, foreground
and host-galaxy extinction (given in Tables \ref{P99a_data} and \ref{P99b_data}), and $\mathrm{ZP}'$ is the zero-point of the
Hubble diagram.

\smallskip
The zero points of the $BVI$ Hubble diagrams derived by \citetalias{phillips99} were 28.671, 28.615, 28.236, respectively.
We note that \citetalias{phillips99} used the \citet{schlegel98} corrections for Galactic reddening, whereas the values in
Tables \ref{P99a_data} and \ref{P99b_data} are in the modern \citet{schlafly11} calibration, which could be a potential
source of systematic error for the derivation of \Ho. However, we checked that this difference has a negligible effect
in our results (0.002 mag difference in $E(B-V)$ for the full sample of 62 SN host galaxies).

\smallskip
Combining the SN peak magnitudes calibrated with TRGB distances with the zero points of the $BVI$ Hubble diagrams
derived by \citetalias{phillips99}, we obtain \Ho\ values of 70.2, 70.1, and 68.7 \kmsmpc\ in $BVI$, respectively,
in good internal agreement given their statistical uncertainty of $\pm$2 \kmsmpc\ (see Table \ref{H0_table}).

\smallskip
Now we apply the \citetalias{phillips99} technique to nearby SNe with Cepheid distances. The resulting values for
\Ho\ range between 72 and 75 \kmsmpc\ (see Table \ref{H0_table}). There is an excellent match with the values obtained
using the \citetalias{hamuy96b} method, thus confirming that the 26 Calán-Tololo SNe were not significantly extinguished
by host-galaxy dust compared to the nearby SNe calibrated with the Cepheid method.

\begin{landscape}
\begin{table}
\caption{Parameters for Individual Supernovae for \citetalias{phillips99} Sample and TRGB distances \label{P99a_data}}
\begin{tabular}{lcccccccccccc}
  \hline
  \multicolumn{1}{l}{SN} &
  \multicolumn{1}{c}{Galaxy} &
  \multicolumn{1}{c}{Distance} &
  \multicolumn{1}{c}{$B_{\mathrm{MAX}}$} &
  \multicolumn{1}{c}{$V_{\mathrm{MAX}}$} &
  \multicolumn{1}{c}{$I_{\mathrm{MAX}}$} &
  \multicolumn{1}{c}{$\Delta m_{15}(B)$} &
  \multicolumn{1}{c}{$E(B-V)_{\mathrm{GAL}}$} &
  \multicolumn{1}{c}{$E(B-V)_{\mathrm{host}}$} &
  \multicolumn{1}{c}{$M^B_{\mathrm{MAX,corr}}$} &
  \multicolumn{1}{c}{$M^V_{\mathrm{MAX,corr}}$} &
  \multicolumn{1}{c}{$M^I_{\mathrm{MAX,corr}}$} &
  \multicolumn{1}{c}{Photometry} \\
  \multicolumn{1}{l}{} &
  \multicolumn{1}{c}{Name} &
  \multicolumn{1}{c}{Modulus} &
  \multicolumn{1}{c}{} &
  \multicolumn{1}{c}{} &
  \multicolumn{1}{c}{} &
  \multicolumn{1}{c}{} &
  \multicolumn{1}{c}{$\pm$0.020} &
  \multicolumn{1}{c}{} &
  \multicolumn{1}{c}{} &
  \multicolumn{1}{c}{} &
  \multicolumn{1}{c}{} &
  \multicolumn{1}{c}{Source} \\
  \multicolumn{1}{l}{} &
  \multicolumn{1}{c}{} &
  \multicolumn{1}{c}{} &
  \multicolumn{1}{c}{} &
  \multicolumn{1}{c}{} &
  \multicolumn{1}{c}{} &
  \multicolumn{1}{c}{} &
  \multicolumn{1}{c}{\citetalias{schlafly11}} &
  \multicolumn{1}{c}{} &
  \multicolumn{1}{c}{} &
  \multicolumn{1}{c}{} &
  \multicolumn{1}{c}{} \\
  \hline
 SN\,1980N  &  N1316   &  31.46(0.04) &  12.49(0.02)  & 12.44(0.02)&  12.71(0.10) &  1.28(0.04) &  0.019 & 0.05(0.02)  & -19.381(0.167)    & -19.339(0.147)   &  -18.932(0.160)  & \citet{hamuy91}     \\
 SN\,1981D  &  N1316   &  31.46(0.04) &  12.59(0.04)  & 12.40(0.04)& --           &  1.44(0.05) &  0.019 & 0.20(0.06)  & -19.980(0.332)    & -19.907(0.283)   &    --            & \citet{hamuy91}     \\
 SN\,1981B  &  N4536   &  30.96(0.05) &  12.03(0.03)  & 11.93(0.03)& --           &  1.10(0.07) &  0.016 & 0.11(0.03)  & -19.462(0.198)    & -19.433(0.170)   &    --            & \citet{buta83}           \\
 SN\,1989B  &  N3627   &  30.22(0.04) &  12.34(0.05)  & 11.99(0.05)&  11.75(0.05) &  1.31(0.07) &  0.029 & 0.34(0.04)  & -19.562(0.252)    & -19.513(0.218)   &  -19.208(0.183)  & \citet{wells94}       \\
 SN\,1998bu &  N3368   &  30.31(0.04) &  12.20(0.03)  & 11.88(0.03)&  11.67(0.05) &  1.01(0.05) &  0.022 & 0.33(0.03)  & -19.521(0.183)    & -19.493(0.153)   &  -19.256(0.126)  & \citet{suntzeff99}    \\
 SN\,1994D  &  N4526   &  31.00(0.07) &  11.89(0.02)  & 11.90(0.02)&  12.06(0.05) &  1.32(0.05) &  0.020 & 0.00(0.02)  & -19.335(0.190)    & -19.280(0.172)   &  -19.039(0.161)  & \citet{richmond95}  \\
 SN\,1994ae &  N3370   &  32.27(0.05) &  13.19(0.02)  & 13.10(0.02)&  13.35(0.03) &  0.86(0.05) &  0.027 & 0.12(0.03)  & -19.479(0.208)    & -19.445(0.182)   &  -19.059(0.156)  & \citet{riess99}     \\
 SN\,1995al &  N3021   &  32.22(0.05) &  13.38(0.02)  & 13.24(0.02)&  13.52(0.05) &  0.83(0.05) &  0.012 & 0.15(0.03)  & -19.271(0.215)    & -19.276(0.189)   &  -18.846(0.170)  & \citet{riess99}     \\
 SN\,2001el &  N1448   &  31.32(0.06) &  12.84(0.02)  & 12.73(0.02)&  12.80(0.04) &  1.15(0.05) &  0.013 & 0.17(0.03)  & -19.289(0.188)    & -19.206(0.159)   &  -18.879(0.131)  & \citet{krisciunas03}\\
 SN\,2002fk &  N1309   &  32.50(0.07) &  13.30(0.03)  & 13.37(0.03)&  13.57(0.03) &  1.02(0.04) &  0.035 & 0.01(0.04)  & -19.321(0.218)    & -19.214(0.180)   &  -18.975(0.137)  & \citet{cartier14}   \\
 SN\,2006dd &  N1316   &  31.46(0.04) &  12.30(0.05)  & 12.27(0.05)&  12.46(0.05) &  1.28(0.05) &  0.019 & 0.07(0.03)  & -19.655(0.202)    & -19.573(0.175)   &  -19.219(0.147)  & \citet{stritzinger10}   \\
 SN\,2007af &  N5584   &  31.82(0.10) &  13.29(0.02)  & 13.25(0.02)& --           &  1.26(0.04) &  0.035 & 0.09(0.05)  & -19.164(0.268)    & -19.058(0.223)   &    --            & \citet{krisciunas17}      \\
 SN\,2011fe &  M101    &  29.08(0.04) &  10.02(0.04)  &  9.96(0.04)&  10.23(0.04) &  1.15(0.04) &  0.008 & 0.09(0.05)  & -19.511(0.245)    & -19.465(0.195)   &  -19.051(0.137)  & \citet{richmond12}  \\
 SN\,2012cg &  N4424   &  31.00(0.06) &  12.11(0.03)  & 11.99(0.03)& --           &  0.92(0.04) &  0.018 & 0.20(0.05)  & -19.648(0.255)    & -19.567(0.208)   &    --            & \citet{marion16}    \\
 SN\,2012fr &  N1365   &  31.36(0.05) &  12.02(0.02)  & 12.04(0.02)& --           &  0.82(0.04) &  0.018 &-0.01(0.11)  & -19.102(0.491)    & -19.106(0.384)   &    --            & \citet{contreras18}      \\
\hline
Mean               &  &              &               &            &              &             &        &             & -19.439           & -19.385          &  -19.050         &                \\
$\sigma$           &  &              &               &            &              &             &        &             &  0.171            &  0.174           &   0.146          &                \\
$\sigma$/$\sqrt n$ &  &              &               &            &              &             &        &             &  0.044            &  0.045           &   0.046          &                \\
Error in Mean      &  &              &               &            &              &             &        &             &  0.057            &  0.048           &   0.047          &                \\
$n$                &  &              &               &            &              &             &        &             &  15               &  15              &   10             &                \\
\hline
\end{tabular}
\end{table}
\end{landscape}

\begin{landscape}
\begin{table}
\caption{Parameters for Individual Supernovae for \citetalias{phillips99} Sample and CEPH distances \label{P99b_data}}
\begin{tabular}{lcccccccccccc}
  \hline
  \multicolumn{1}{l}{SN} &
  \multicolumn{1}{c}{Galaxy} &
  \multicolumn{1}{c}{Distance} &
  \multicolumn{1}{c}{$B_{\mathrm{MAX}}$} &
  \multicolumn{1}{c}{$V_{\mathrm{MAX}}$} &
  \multicolumn{1}{c}{$I_{\mathrm{MAX}}$} &
  \multicolumn{1}{c}{$\Delta m_{15}(B)$} &
  \multicolumn{1}{c}{$E(B-V)_{\mathrm{GAL}}$} &
  \multicolumn{1}{c}{$E(B-V)_{\mathrm{host}}$} &
  \multicolumn{1}{c}{$M^B_{\mathrm{MAX,corr}}$} &
  \multicolumn{1}{c}{$M^V_{\mathrm{MAX,corr}}$} &
  \multicolumn{1}{c}{$M^I_{\mathrm{MAX,corr}}$} &
  \multicolumn{1}{c}{Photometry} \\
  \multicolumn{1}{l}{} &
  \multicolumn{1}{c}{Name} &
  \multicolumn{1}{c}{Modulus} &
  \multicolumn{1}{c}{} &
  \multicolumn{1}{c}{} &
  \multicolumn{1}{c}{} &
  \multicolumn{1}{c}{} &
  \multicolumn{1}{c}{$\pm$0.020} &
  \multicolumn{1}{c}{} &
  \multicolumn{1}{c}{} &
  \multicolumn{1}{c}{} &
  \multicolumn{1}{c}{} &
  \multicolumn{1}{c}{Source} \\
  \multicolumn{1}{l}{} &
  \multicolumn{1}{c}{} &
  \multicolumn{1}{c}{} &
  \multicolumn{1}{c}{} &
  \multicolumn{1}{c}{} &
  \multicolumn{1}{c}{} &
  \multicolumn{1}{c}{} &
  \multicolumn{1}{c}{\citetalias{schlafly11}} &
  \multicolumn{1}{c}{} &
  \multicolumn{1}{c}{} &
  \multicolumn{1}{c}{} &
  \multicolumn{1}{c}{} \\
  \hline
 SN\,1981B  &  N4536   & 30.877(0.053)&  12.03(0.03)  & 11.93(0.03)& --           &  1.10(0.07) &  0.016 & 0.11(0.03)        & -19.379(0.199)    & -19.350(0.171)   & --               & \citet{buta83} \\
 SN\,1990N  &  N4639   & 31.503(0.071)&  12.76(0.03)  & 12.70(0.02)&  12.94(0.02) &  1.07(0.05) &  0.023 & 0.09(0.03)        & -19.196(0.191)    & -19.143(0.161)   &  -18.760(0.129)  & \citet{lira98}   \\
 SN\,1994ae &  N3370   & 32.043(0.049)&  13.19(0.02)  & 13.10(0.02)&  13.35(0.03) &  0.86(0.05) &  0.027 & 0.12(0.03)        & -19.252(0.208)    & -19.218(0.181)   &  -18.832(0.156)  & \citet{riess99}     \\
 SN\,1995al &  N3021   & 32.469(0.090)&  13.38(0.02)  & 13.24(0.02)&  13.52(0.05) &  0.83(0.05) &  0.012 & 0.15(0.03)        & -19.520(0.227)    & -19.525(0.204)   &  -19.095(0.186)  & \citet{riess99}     \\
 SN\,1998aq &  N3982   & 31.708(0.069)&  12.35(0.02)  & 12.46(0.02)&  12.69(0.02) &  1.09(0.04) &  0.012 & 0.02(0.06)        & -19.485(0.284)    & -19.343(0.224)   &  -19.073(0.154)  & \citet{riess05}  \\
 SN\,2001el &  N1448   & 31.282(0.045)&  12.84(0.02)  & 12.73(0.02)&  12.80(0.04) &  1.15(0.05) &  0.013 & 0.17(0.03)        & -19.251(0.184)    & -19.168(0.154)   &  -18.841(0.125)  & \citet{krisciunas03}\\
 SN\,2002fk &  N1309   & 32.494(0.055)&  13.30(0.03)  & 13.37(0.03)&  13.57(0.03) &  1.02(0.04) &  0.035 & 0.01(0.04)        & -19.315(0.214)    & -19.208(0.174)   &  -18.969(0.130)  & \citet{cartier14}   \\
 SN\,2003du &  U9391   & 32.890(0.063)&  13.44(0.04)  & 13.55(0.04)&  13.84(0.02) &  1.09(0.05) &  0.009 & 0.01(0.05)        & -19.522(0.253)    & -19.394(0.204)   &  -19.081(0.145)  & \citet{hicken09} \\
 SN\,2005cf &  N5917   & 32.234(0.102)&  13.63(0.02)  & 13.56(0.02)& --           &  1.03(0.04) &  0.086 & 0.06(0.09)        & -19.158(0.406)    & -19.087(0.318)   & --               & \citet{hicken09} \\
 SN\,2007af &  N5584   & 31.757(0.046)&  13.29(0.02)  & 13.25(0.02)& --           &  1.26(0.04) &  0.035 & 0.09(0.05)        & -19.101(0.252)    & -18.995(0.205)   & --               & \citet{krisciunas17} \\
 SN\,2009ig &  N1015   & 32.468(0.081)&  13.58(0.04)  & 13.46(0.02)& --           &  0.86(0.04) &  0.029 & 0.10(0.12)        & -19.210(0.529)    & -19.225(0.410)   & --               & \citet{hicken12} \\
 SN\,2011by &  N3972   & 31.558(0.070)&  12.94(0.02)  & 12.89(0.02)&  12.97(0.02) &  1.07(0.04) &  0.013 & 0.13(0.04)        & -19.198(0.214)    & -19.105(0.176)   &  -18.841(0.132)  & \citet{stahl19} \\
 SN\,2011fe &  M101    & 29.106(0.045)&  10.02(0.04)  &  9.96(0.04)&  10.23(0.04) &  1.15(0.04) &  0.008 & 0.09(0.05)        & -19.537(0.246)    & -19.491(0.196)   &  -19.077(0.138)  & \citet{richmond12}  \\
 SN\,2012cg &  N4424   & 31.051(0.292)&  12.11(0.03)  & 11.99(0.03)& --           &  0.92(0.04) &  0.018 & 0.19(0.05)        & -19.661(0.383)    & -19.589(0.354)   & --               & \citet{marion16}    \\
 SN\,2012fr &  N1365   & 31.278(0.057)&  12.02(0.02)  & 12.04(0.02)& --           &  0.82(0.04) &  0.018 &-0.01(0.11)        & -19.020(0.492)    & -19.024(0.385)   & --               & \citet{contreras18} \\
 SN\,2012ht &  N3447   & 31.879(0.043)&  13.11(0.02)  & 13.10(0.02)& --           &  1.29(0.04) &  0.026 & 0.01(0.03)        & -19.046(0.193)    & -18.997(0.165)   & --               & \citet{burns18}  \\
 SN\,2013dy &  N7250   & 31.470(0.078)&  13.30(0.02)  & 12.95(0.02)&  12.95(0.02) &  0.84(0.04) &  0.135 & 0.20(0.10)        & -19.336(0.451)    & -19.373(0.354)   &  -18.992(0.239)  & \citet{pan15}        \\
 SN\,2015F  &  N2442   & 31.482(0.053)&  13.49(0.02)  & 13.29(0.02)& --           &  1.43(0.04) &  0.180 & 0.09(0.05)        & -19.308(0.291)    & -19.195(0.251)   & --               & \citet{burns18}  \\
\hline
Mean               &  &              &               &            &              &             &        &                   & -19.297           & -19.229          &  -18.938         &                \\
$\sigma$           &  &              &               &            &              &             &        &                   &  0.166            &  0.168           &   0.129          &                \\
$\sigma$/$\sqrt n$ &  &              &               &            &              &             &        &                   &  0.039            &  0.040           &   0.041          &                \\
Error in Mean      &  &              &               &            &              &             &        &                   &  0.058            &  0.049           &   0.046          &                \\
$n$                &  &              &               &            &              &             &        &                   &  18               &  18              &   10             &                \\
\hline
\end{tabular}
\end{table}
\end{landscape}

\subsection{The F10 METHODOLOGY} \label{F10}

The decade of the 90s meant a breakthrough for the measurement of the expansion rate of the Universe using SNe\,Ia,
thanks to the gathering of digital CCD photometry of several dozens of SNe in the Hubble flow. However, the analysis
of such data promptly revealed that the transformation of the instrumental magnitudes to the standard photometric system
was rendered challenging owing to the non-stellar nature of the SN spectral energy distributions. Differences of several
hundreds of a magnitude were noticed in the light curves of the same objects observed with different instruments
\citep{suntzeff00,stritzinger02}. An additional difficulty in the standardization of SNe~Ia as distance indicators arose
from the effects of dust extinction in the SN parent galaxies which, despite the efforts to determine them from the observed
SN colours, introduced significant uncertainties more strongly on the bluer wavelengths. These problems were addressed by
the Carnegie Supernova Program (CSP) launched in 2004 \citep{hamuy06} from Las Campanas Observatory (LCO) which, after nearly
a decade of effort, led to the gathering of high-quality optical/NIR ($uBgVriYJHK$) lightcurves of 134 SNe Ia light curves
in the Hubble flow with stable instrumental systems, namely, the Swope 1m and the du Pont 2.5m telescopes.

\smallskip
\citet{contreras10} published the first data release (DR1) of 34 SN light curves observed between 2004-2006. Since the
observations were consistently obtained with the same instrumental bandpasses, the instrumental magnitudes were converted
to the natural system through the application of a zero point and no colour term, thus avoiding the difficulty of transforming
the data to the standard photometric system.

\smallskip
Following on the approach of \citetalias{phillips99}, this high-quality dataset allowed \citetalias{folatelli10} to derive
an improved derivation of the ``Lira law'', as well as better relationships between near-maximum reddening-free colours and
\dm, with a precision between 0.06-0.14 mag (see their table 3). Each of these ten calibrations allowed them to derive precise
colour-excesses and study in depth the reddening law caused by host-galaxy dust.

\smallskip
The colour excesses were then used to re-examine the correlation of reddening-corrected absolute peak magnitudes versus
decline rate, in the same manner as in \citetalias{phillips99}, and re-assess the precision to which SNe Ia could be used
as standardizable candles. As shown in their equation (7) the following two-parameter model was adopted:

\begin{equation}
\tilde \mu = m_X - M_X(0) - b_X [\Delta m_{15}(B) - 1.1] - R_X^{YZ} E(Y-Z)
\end{equation}

\smallskip
\noindent where the three measured variables are $m_X$, the peak apparent magnitude of the SN in a given band X corrected
for K terms and Galactic reddening \citep{schlegel98}, $E(Y-Z)$, the colour excess due to host-galaxy dust obtained from bands
Y and Z, the decline rate \dm, and the distance modulus $\tilde\mu$ derived from the host-galaxy redshift and the cosmological
parameters $\Omega_{\Lambda}$=0.72, $\Omega_{M}$=0.28, and \Ho=72 (see their equation 5). In this model there are three fitting
parameters: the slope of the luminosity versus decline-rate, $b_X$, the slope of the luminosity versus colour-excess, $R_X^{YZ}$,
and the peak absolute magnitude of the SNe~Ia with \dm=1.1 and zero colour-excess, $M_X(0)$. Their Table 5 shows results of the
fits for ten (X,Y,Z) combinations, from the 23 ``Best-observed'' SN subsample, which are characterized by rms dispersions between
0.12-0.15 mag. These fits are restricted to the range 0.7 \textless \dm \textless 1.7 over which the colour excess calibrations are valid.

\subsubsection{Absolute Magnitudes} \label{F10_AM}

Table \ref{F10a_data} summarizes the input parameters for the six nearby SNe having TRGB distances and for which we are able to
apply the \citetalias{folatelli10} technique, that is, SNe with (1) NIR photometry available in the natural CSP system and
(2) having decline rates within the range 0.7 \textless \dm \textless 1.7. Two of these six SNe were observed by the CSP
(SN\,2007af and SN\,2012fr), two were observed with other instruments but were transformed to the Swope system via S-corrections
(SN\,2001el and SN\,2006dd), one was observed with the FLWO/PAIRITEL instrument and converted from the 2MASS to the CSP system
using the offsets determined by \citet{contreras10} (SN\,2012cg), and one observed with the LCO du Pont WIRC instrument (SN\,2002fk)
which is virtually identical to the CSP photometric system. We measure peak magnitudes, decline rates, and colour excesses directly
from the light curves, from which we compute standardized absolute peak magnitudes as follows,

\begin{equation}
M^{\mathrm{corr}}_{X} = m_{X} - A_{\mathrm{GAL}} - b_X [\Delta m_{15}(B) - 1.1] - R_X^{YZ} E(Y-Z) - \mu
\end{equation}

\smallskip
\noindent where $\mu$ is the TRGB distance modulus. The resulting values are given in Table \ref{F10a_data}. The mean absolute magnitudes are shown at the bottom of Table \ref{F10a_data} for
the $J$ and $H$ bands (we omit the results for the remaining bands which only have two SNe calibrated with the TRGB method). The nearby
SNe yield a dispersion in the standardized absolute magnitudes of 0.12 and 0.14 mag in $J$ and $H$, respectively, in good agreement
with the expected values yielded by the distant sample.

\smallskip
Now we apply the same technique for the nine SNe having Cepheid distances and $J$ and $H$ photometry. Four of these nine SNe were
observed by the CSP (SN\,2007af, SN\,2012fr, SN\,2012ht, and SN\,2015F), one was observed with other instruments but was transformed to
the Swope system via S-corrections (SN\,2001el), three were observed with the FLWO/PAIRITEL instrument and converted from the 2MASS to
the CSP system using the offsets determined by \citet{contreras10} (SN\,2005cf, SN\,2011by, and SN\,2012cg), and one was observed with
the LCO du Pont WIRC instrument which is virtually identical to the CSP photometric system (SN\,2002fk). Table \ref{F10b_data} presents
the relevant parameters of the SNe along with the distance moduli published by \citetalias{riess16} to which we add a global correction
of -0.029 mag (= 5 log 73.24/74.22) in order to place them in the \citetalias{riess19} Cepheid scale.

\subsubsection{The Hubble-Lema\^itre constant} \label{F10_H0}

Armed with the SN standardized peak magnitudes we turn now to the determination of the value of \Ho\ by means of the following expression:

\begin{equation}
H_0 (X) = 72 \times 10^{0.2[M^{\mathrm{corr}}_{X}-M_X(0)]}
\end{equation}

\smallskip
\noindent where $M^{\mathrm{corr}}_{X}$ is the mean absolute magnitude in a given band X corrected for foreground extinction, decline rate,
and colour excess, derived from the nearby SNe, while $M_X(0)$ is the standardized peak absolute magnitude derived by \citetalias{folatelli10}
from the distant SNe~Ia, namely -18.44$\pm$0.01 and -18.38$\pm$0.02 in $J$ and $H$, respectively. The resulting values using the TRGB distance moduli are \Ho(J)=66.5$\pm$1.6 and \Ho(H)=69.4$\pm$2.1 \kmsmpc (see Table \ref{H0_table}).
Adopting the Cepheid distances, we obtain \Ho(J)=69.1$\pm$1.3 and \Ho(H)=74.4$\pm$1.6 \kmsmpc,
respectively (see Table \ref{H0_table}). We note that there is a 2.6$\sigma$ difference between both values. As shown in the next section the updated NIR CSP calibration by \citetalias{kattner12} significantly alleviates this tension between
the $J$ and $H$ bands.

\begin{landscape}
\begin{table}
\caption{Parameters for Individual Supernovae for \citetalias{folatelli10} Sample and TRGB distances \label{F10a_data}}
\begin{tabular}{lcccccccccc}
  \hline
  \multicolumn{1}{l}{SN} &
  \multicolumn{1}{c}{Galaxy} &
  \multicolumn{1}{c}{Distance} &
  \multicolumn{1}{c}{$V_{\mathrm{MAX}}$} &
  \multicolumn{1}{c}{$J_{\mathrm{MAX}}$} &
  \multicolumn{1}{c}{$H_{\mathrm{MAX}}$} &
  \multicolumn{1}{c}{$\Delta m_{15}(B)$} &
  \multicolumn{1}{c}{$E(B-V)_{\mathrm{GAL}}$} &
  \multicolumn{1}{c}{$M_J$$^{\mathrm{corr}}$} &
  \multicolumn{1}{c}{$M_H$$^{\mathrm{corr}}$} &
  \multicolumn{1}{c}{Photometry} \\
  \multicolumn{1}{l}{} &
  \multicolumn{1}{c}{Name} &
  \multicolumn{1}{c}{Modulus} &
  \multicolumn{1}{c}{} &
  \multicolumn{1}{c}{} &
  \multicolumn{1}{c}{} &
  \multicolumn{1}{c}{} &
  \multicolumn{1}{c}{$\pm$0.020} &
  \multicolumn{1}{c}{} &
  \multicolumn{1}{c}{} &
  \multicolumn{1}{c}{Source} \\
  \multicolumn{1}{l}{} &
  \multicolumn{1}{c}{} &
  \multicolumn{1}{c}{} &
  \multicolumn{1}{c}{} &
  \multicolumn{1}{c}{} &
  \multicolumn{1}{c}{} &
  \multicolumn{1}{c}{} &
  \multicolumn{1}{c}{\citetalias{schlafly11}} &
  \multicolumn{1}{c}{} &
  \multicolumn{1}{c}{} &
  \multicolumn{1}{c}{} \\
  \hline
 SN\,2001el &  N1448   &  31.32(0.06) &  12.73(0.02)  & 12.90(0.04)&  13.08(0.04) &   1.15(0.05) &  0.014 &    -18.517(0.084) & -18.335(0.084) &  \citet{krisciunas03}  \\
 SN\,2002fk &  N1309   &  32.50(0.07) &  13.37(0.03)  & 13.76(0.02)&  13.98(0.02) &   1.02(0.04) &  0.038 &    -18.755(0.080) & -18.537(0.079) &  \citet{cartier14}     \\
 SN\,2006dd &  N1316   &  31.46(0.04) &  12.31(0.02)  & 12.73(0.05)&  12.84(0.05) &   1.08(0.04) &  0.020 &    -18.761(0.072) & -18.665(0.071) &  \citet{stritzinger10} \\
 SN\,2007af &  N5584   &  31.82(0.10) &  13.18(0.03)  & 13.45(0.02)&  13.62(0.02) &   1.23(0.04) &  0.038 &    -18.507(0.108) & -18.295(0.108) &  \citet{krisciunas17}  \\
 SN\,2012cg &  N4424   &  31.00(0.06) &  11.99(0.03)  & 12.34(0.05)&  12.52(0.04) &   0.92(0.04) &  0.019 &    -18.619(0.088) & -18.491(0.087) &  \citet{marion16}      \\
 SN\,2012fr &  N1365   &  31.36(0.05) &  11.99(0.02)  & 12.73(0.02)&  12.96(0.02) &   0.83(0.04) &  0.020 &    -18.490(0.067) & -18.304(0.077) &  \citet{contreras18}   \\
\hline
Mean               &  &              &               &            &              &             &         &    -18.614        & -18.459        &                  \\
$\sigma$           &  &              &               &            &              &             &         &      0.129        &  0.156         &                  \\
$\sigma$/$\sqrt n$ &  &              &               &            &              &             &         &      0.053        &  0.064         &                  \\
Error in Mean      &  &              &               &            &              &             &         &      0.033        &  0.034         &                  \\
$n$                &  &              &               &            &              &             &         &      6            &  6             &                  \\
\hline
\end{tabular}
\end{table}
\end{landscape}

\begin{landscape}
\begin{table}
\caption{Parameters for Individual Supernovae for \citetalias{folatelli10} Sample and CEPH distances \label{F10b_data}}
\begin{tabular}{lcccccccccc}
  \hline
  \multicolumn{1}{l}{SN} &
  \multicolumn{1}{c}{Galaxy} &
  \multicolumn{1}{c}{Distance} &
  \multicolumn{1}{c}{$V_{\mathrm{MAX}}$} &
  \multicolumn{1}{c}{$J_{\mathrm{MAX}}$} &
  \multicolumn{1}{c}{$H_{\mathrm{MAX}}$} &
  \multicolumn{1}{c}{$\Delta m_{15}(B)$} &
  \multicolumn{1}{c}{$E(B-V)_{\mathrm{GAL}}$} &
  \multicolumn{1}{c}{$M_J$$^{\mathrm{corr}}$} &
  \multicolumn{1}{c}{$M_H$$^{\mathrm{corr}}$} &
  \multicolumn{1}{c}{Photometry} \\
  \multicolumn{1}{l}{} &
  \multicolumn{1}{c}{Name} &
  \multicolumn{1}{c}{Modulus} &
  \multicolumn{1}{c}{} &
  \multicolumn{1}{c}{} &
  \multicolumn{1}{c}{} &
  \multicolumn{1}{c}{} &
  \multicolumn{1}{c}{$\pm$0.020} &
  \multicolumn{1}{c}{} &
  \multicolumn{1}{c}{} &
  \multicolumn{1}{c}{Source} \\
  \multicolumn{1}{l}{} &
  \multicolumn{1}{c}{} &
  \multicolumn{1}{c}{} &
  \multicolumn{1}{c}{} &
  \multicolumn{1}{c}{} &
  \multicolumn{1}{c}{} &
  \multicolumn{1}{c}{} &
  \multicolumn{1}{c}{\citetalias{schlafly11}} &
  \multicolumn{1}{c}{} &
  \multicolumn{1}{c}{} &
  \multicolumn{1}{c}{} \\
  \hline
 SN\,2001el &  N1448   &  31.282(0.045) &  12.73(0.02)  & 12.90(0.04)&  13.08(0.04) &   1.15(0.05) &  0.014 &    -18.479(0.074) & -18.297(0.074) &  \citet{krisciunas03} \\
 SN\,2002fk &  N1309   &  32.494(0.055) &  13.37(0.03)  & 13.76(0.02)&  13.98(0.02) &   1.02(0.04) &  0.038 &    -18.749(0.068) & -18.531(0.066) &  \citet{cartier14} \\
 SN\,2005cf &  N5917   &  32.234(0.102) &  13.56(0.02)  & 13.82(0.05)&  13.95(0.05) &   1.01(0.04) &  0.086 &    -18.469(0.119) & -18.345(0.119) &  \citet{hicken09} \\
 SN\,2007af &  N5584   &  31.757(0.046) &  13.18(0.03)  & 13.45(0.02)&  13.62(0.02) &   1.23(0.04) &  0.038 &    -18.444(0.062) & -18.232(0.062) &  \citet{krisciunas17}  \\
 SN\,2011by &  N3972   &  31.558(0.070) &  12.89(0.03)  & 13.19(0.02)&  13.48(0.05) &   1.13(0.04) &  0.013 &    -18.437(0.081) & -18.121(0.091) &  \citet{friedman15,stahl19} \\
 SN\,2012cg &  N4424   &  31.051(0.292) &  11.99(0.03)  & 12.34(0.05)&  12.52(0.04) &   0.92(0.04) &  0.019 &    -18.670(0.299) & -18.542(0.299) &  \citet{marion16} \\
 SN\,2012fr &  N1365   &  31.278(0.057) &  11.99(0.02)  & 12.73(0.02)&  12.96(0.02) &   0.83(0.04) &  0.020 &    -18.408(0.072) & -18.222(0.082) &  \citet{contreras18} \\
 SN\,2012ht &  N3447   &  31.879(0.043) &  13.06(0.02)  & 13.45(0.02)&  13.62(0.02) &   1.25(0.04) &  0.026 &    -18.553(0.059) & -18.334(0.060) &  \citet{burns18}  \\
 SN\,2015F  &  N2442   &  31.482(0.053) &  13.26(0.02)  & 13.12(0.05)&  13.42(0.10) &   1.23(0.04) &  0.180 &    -18.637(0.082) & -18.222(0.120) &  \citet{burns18}  \\
\hline
Mean               &  &              &               &            &              &             &         &    -18.529        & -18.309        &                  \\
$\sigma$           &  &              &               &            &              &             &         &      0.119        &  0.127         &                  \\
$\sigma$/$\sqrt n$ &  &              &               &            &              &             &         &      0.040        &  0.042         &                  \\
Error in Mean      &  &              &               &            &              &             &         &      0.026        &  0.027         &                  \\
$n$                &  &              &               &            &              &             &         &      9            &  9             &                  \\
\hline
\end{tabular}
\end{table}
\end{landscape}

\subsection{THE K12 METHODOLOGY} \label{K12}

\citet{kattner12} reanalized the standardization of SNe\,Ia in the NIR, in a similar manner as \citetalias{folatelli10},
but limiting the CSP sample to the 27 best-observed SNe, namely, those having pre-maximum coverage in optical bands
and, particularly the sub-sample of 13 objects also having having pre-maximum NIR observations. The latter condition is
particularly relevant since, as shown by \citetalias{folatelli10}, the extrapolation of peak magnitudes using NIR template
light curves could introduce significant errors.

\smallskip
The correlation between peak absolute luminosity and decline rate was investigated using the same equation first proposed
by \citetalias{phillips99},

\begin{equation}
\tilde \mu = m_X - M_X(0) - b_X [\Delta m_{15}(B) - 1.1] - R_X E(B-V)
\end{equation}

\smallskip
\noindent where the measured quantities are $m_X$, the peak apparent magnitude of the SN in a given band X (X=Y,J,H)
corrected for K terms and Galactic reddening \citep{schlegel98}, \dm, the decline rate measured from the $B$-band, and
$E(B-V)$, the colour excess due to host-galaxy reddening derived from the near-maximum reddening-free
$B_{\mathrm{MAX}}$-$V_{\mathrm{MAX}}$ colour derived by \citetalias{folatelli10}. As in \citetalias{folatelli10},
the left-hand term of this equation is the distance modulus $\tilde\mu$ derived from the host-galaxy redshift and the
cosmological parameters $\Omega_{\Lambda}$=0.72, $\Omega_{M}$=0.28, and \Ho=72. In this model $R_X$
is the total-to-selective absorption coefficient for band X, a fixed parameter of $R_Y$=1.18, $R_J$=0.89, $R_H$=0.57,
for an adopted $R_V$=3.1 dust extinction law. In this model there are two fitting parameters: the slope of the
luminosity versus decline-rate relation, $b_X$, and the peak absolute magnitude of the SNe Ia with \dm=1.1
and zero colour-excess, $M_X(0)$.

\smallskip
Their Table 5 shows results of the fits for the three NIR bands ($Y$,$J$,$H$) and five different sub-samples of SNe.
Here we use sub-sample 3, which uses SNe~Ia with first observations starting within five days after NIR peak
brightness and excludes the highly reddened and fast-declining events. The correlations are characterized by rms
dispersions between 0.09-0.12 mag. These fits are restricted to the range 0.7 \textless \dm \textless 1.7 over which
the colour excess calibration is valid.

\subsubsection{Absolute Magnitudes} \label{K12_AM}

We measure peak magnitudes, decline rates and colour excesses for the six nearby SNe having TRGB distances and
for which we are able to apply the \citetalias{kattner12} technique, that is, SNe with (1) NIR photometry available
in the natural CSP system and (2) having decline rates within the range 0.7 \textless \dm \textless 1.7. We compute standardized absolute peak magnitudes as follows,

\begin{equation}
M^{\mathrm{corr}}_{X} = m_{X} - A_{\mathrm{GAL}} - b_X [\Delta m_{15}(B) - 1.1] - R_X^{YZ} E(Y-Z) - \mu
\end{equation}

\smallskip
\noindent where $\mu$ is the TRGB distance modulus. Table \ref{K12a_data} summarizes the input parameters
and their standardized absolute peak magnitudes for all six SNe. The mean value is shown at the bottom of
Table \ref{K12a_data} for the $J$ and $H$ different bands (we omit the $Y$-band as there are only two nearby
SNe with TRGB distance). The nearby SNe yield dispersions in the corrected absolute magnitudes of 0.08 and
0.11 mag, similar to those obtained by the distant sample.

\smallskip
Now we apply the \citetalias{kattner12} technique to the sample of nine nearby SNe with Cepheid distances and NIR photometry in the CSP natural system, and decline rates within the range 0.7 \textless \dm \textless 1.7.
Table \ref{K12b_data} presents the relevant parameters of the SNe along with the distance moduli published
by \citetalias{riess16} to which we add a global correction of -0.029 mag (= 5 log 73.24/74.22) in order
to place them in the \citetalias{riess19} Cepheid scale, and their corresponding standardized absolute peak
magnitudes. 

\subsubsection{The Hubble-Lema\^itre constant} \label{K12_H0}

As in \citetalias{folatelli10}, the value of the Hubble-Lema\^itre constant can be calculated using the following
expression:

\begin{equation}
H_0 (X) = 72 \times 10^{0.2[M^{\mathrm{corr}}_{X}-M_X(0)]}
\end{equation}

\smallskip
\noindent where $M^{\mathrm{corr}}_{X}$ is the mean standardized absolute peak magnitude in a given band X corrected
for foreground extinction, decline rate, and colour excess, derived from the nearby SNe, while $M_X(0)$ is
the standardized peak absolute magnitude derived by \citetalias{kattner12} from the distant SNe\,Ia,
namely -18.552$\pm$0.002 and -18.390$\pm$0.003 in $J$ and $H$, respectively.

\smallskip
As can be seen in Table \ref{H0_table}, the values for \Ho\ obtained for $J$ and $H$ bands are 69.2$\pm$1.2
and 70.3$\pm$1.6, respectively. 
Adopting the Cepheid distances, the resulting values for \Ho\ from the $J$ and $H$ bands are 72.7$\pm$1.0
and 75.2$\pm$1.2 \kmsmpc, respectively. 
As anticipated in the previous section, the \citetalias{kattner12} recalibration of the $J$-band SN Ia
luminosity clearly alleviates the tension between the $J$ and $H$ bands calibration derived from
\citetalias{folatelli10}.

\begin{landscape}
\begin{table}
\caption{Parameters for Individual Supernovae for \citetalias{kattner12} Sample and TRGB distances \label{K12a_data}}
\begin{tabular}{lccccccccccc}
\hline
\multicolumn{1}{l}{SN} &
\multicolumn{1}{c}{Galaxy} &
\multicolumn{1}{c}{Distance} &
\multicolumn{1}{c}{$B_{\mathrm{MAX}}$} &
\multicolumn{1}{c}{$V_{\mathrm{MAX}}$} &
\multicolumn{1}{c}{$J_{\mathrm{MAX}}$} &
\multicolumn{1}{c}{$H_{\mathrm{MAX}}$} &
\multicolumn{1}{c}{$\Delta m_{15}(B)$} &
\multicolumn{1}{c}{$E(B-V)_{\mathrm{GAL}}$} &
\multicolumn{1}{c}{$M_J$$^{\mathrm{corr}}$} &
\multicolumn{1}{c}{$M_H$$^{\mathrm{corr}}$} &
\multicolumn{1}{c}{Photometry} \\
\multicolumn{1}{l}{} &
\multicolumn{1}{c}{Name} &
\multicolumn{1}{c}{Modulus} &
\multicolumn{1}{c}{} &
\multicolumn{1}{c}{} &
\multicolumn{1}{c}{} &
\multicolumn{1}{c}{} &
\multicolumn{1}{c}{} &
\multicolumn{1}{c}{$\pm$0.020} &
\multicolumn{1}{c}{} &
\multicolumn{1}{c}{} &
\multicolumn{1}{c}{Source}\\
  \multicolumn{1}{l}{} &
  \multicolumn{1}{c}{} &
  \multicolumn{1}{c}{} &
  \multicolumn{1}{c}{} &
  \multicolumn{1}{c}{} &
  \multicolumn{1}{c}{} &
  \multicolumn{1}{c}{} &
  \multicolumn{1}{c}{} &
  \multicolumn{1}{c}{\citetalias{schlafly11}} &
  \multicolumn{1}{c}{} &
  \multicolumn{1}{c}{} \\
\hline
 SN\,2001el &  N1448   &  31.32(0.06) &  12.84(0.02)  & 12.73(0.02)&  12.90(0.04)&  13.08(0.04) &  1.15(0.05) &  0.014 &  -18.550(0.124) & -18.326(0.097) & \citet{krisciunas03}  \\
 SN\,2002fk &  N1309   &  32.50(0.07) &  13.30(0.03)  & 13.37(0.03)&  13.76(0.02)&  13.98(0.02) &  1.02(0.04) &  0.038 &  -18.673(0.127) & -18.473(0.099) & \citet{cartier14}     \\
 SN\,2006dd &  N1316   &  31.46(0.04) &  12.30(0.05)  & 12.31(0.02)&  12.73(0.05)&  12.84(0.05) &  1.08(0.04) &  0.020 &  -18.733(0.126) & -18.621(0.095) & \citet{stritzinger10} \\
 SN\,2007af &  N5584   &  31.82(0.10) &  13.28(0.03)  & 13.18(0.03)&  13.45(0.02)&  13.62(0.02) &  1.23(0.04) &  0.038 &  -18.514(0.147) & -18.299(0.123) & \citet{krisciunas17}  \\
 SN\,2012cg &  N4424   &  31.00(0.06) &  12.11(0.03)  & 11.99(0.03)&  12.34(0.05)&  12.52(0.04) &  0.92(0.04) &  0.019 &  -18.733(0.133) & -18.518(0.101) & \citet{marion16}      \\
 SN\,2012fr &  N1365   &  31.36(0.05) &  12.03(0.02)  & 11.99(0.02)&  12.73(0.02)&  12.96(0.02) &  0.83(0.04) &  0.020 &  -18.606(0.120) & -18.371(0.090) & \citet{contreras18}   \\
\hline
Mean               &  &              &               &            &              &             &              &        &  -18.638         &  -18.442       &                \\
$\sigma$           &  &              &               &            &              &             &              &        &    0.090         &   0.123        &                \\
$\sigma$/$\sqrt n$ &  &              &               &            &              &             &              &        &    0.037         &   0.050        &                \\
Error in Mean      &  &              &               &            &              &             &              &        &    0.053         &   0.041        &                \\
$n$                &  &              &               &            &              &             &              &        &    6             &   6            &                \\
\hline
\end{tabular}
\end{table}
\end{landscape}

\begin{landscape}
\begin{table}
\caption{Parameters for Individual Supernovae for \citetalias{kattner12} Sample and CEPH distances \label{K12b_data}}
\begin{tabular}{lccccccccccc}
\hline
\multicolumn{1}{l}{SN} &
\multicolumn{1}{c}{Galaxy} &
\multicolumn{1}{c}{Distance} &
\multicolumn{1}{c}{$B_{\mathrm{MAX}}$} &
\multicolumn{1}{c}{$V_{\mathrm{MAX}}$} &
\multicolumn{1}{c}{$J_{\mathrm{MAX}}$} &
\multicolumn{1}{c}{$H_{\mathrm{MAX}}$} &
\multicolumn{1}{c}{$\Delta m_{15}(B)$} &
\multicolumn{1}{c}{$E(B-V)_{\mathrm{GAL}}$} &
\multicolumn{1}{c}{$M_J$$^{\mathrm{corr}}$} &
\multicolumn{1}{c}{$M_H$$^{\mathrm{corr}}$} &
\multicolumn{1}{c}{Photometry} \\
\multicolumn{1}{l}{} &
\multicolumn{1}{c}{Name} &
\multicolumn{1}{c}{Modulus} &
\multicolumn{1}{c}{} &
\multicolumn{1}{c}{} &
\multicolumn{1}{c}{} &
\multicolumn{1}{c}{} &
\multicolumn{1}{c}{} &
\multicolumn{1}{c}{$\pm$0.020} &
\multicolumn{1}{c}{} &
\multicolumn{1}{c}{} &
\multicolumn{1}{c}{Source} \\
  \multicolumn{1}{l}{} &
  \multicolumn{1}{c}{} &
  \multicolumn{1}{c}{} &
  \multicolumn{1}{c}{} &
  \multicolumn{1}{c}{} &
  \multicolumn{1}{c}{} &
  \multicolumn{1}{c}{} &
  \multicolumn{1}{c}{} &
  \multicolumn{1}{c}{\citetalias{schlafly11}} &
  \multicolumn{1}{c}{} &
  \multicolumn{1}{c}{} \\
\hline
 SN\,2001el &  N1448   &  31.282(0.045) &  12.84(0.02)  & 12.73(0.02)&  12.90(0.04)&  13.08(0.04) &  1.15(0.05) &  0.014 &  -18.512(0.118) & -18.288(0.089) & \citet{krisciunas03} \\
 SN\,2002fk &  N1309   &  32.494(0.055) &  13.30(0.03)  & 13.37(0.03)&  13.76(0.02)&  13.98(0.02) &  1.02(0.04) &  0.038 &  -18.667(0.120) & -18.467(0.090) & \citet{cartier14} \\
 SN\,2005cf &  N5917   &  32.234(0.102) &  13.62(0.02)  & 13.56(0.02)&  13.82(0.05)&  13.95(0.05) &  1.01(0.04) &  0.086 &  -18.460(0.152) & -18.309(0.131) & \citet{hicken09} \\
 SN\,2007af &  N5584   &  31.757(0.046) &  13.28(0.03)  & 13.18(0.03)&  13.45(0.02)&  13.62(0.02) &  1.23(0.04) &  0.038 &  -18.451(0.117) & -18.236(0.085) & \citet{krisciunas17} \\
 SN\,2011by &  N3972   &  31.558(0.070) &  12.94(0.03)  & 12.89(0.03)&  13.19(0.02)&  13.48(0.05) &  1.13(0.04) &  0.013 &  -18.439(0.127) & -18.125(0.109) & \citet{friedman15,stahl19} \\
 SN\,2012cg &  N4424   &  31.051(0.292) &  12.11(0.03)  & 11.99(0.03)&  12.34(0.05)&  12.52(0.04) &  0.92(0.04) &  0.019 &  -18.784(0.315) & -18.569(0.303) & \citet{marion16} \\
 SN\,2012fr &  N1365   &  31.278(0.057) &  12.03(0.02)  & 11.99(0.02)&  12.73(0.02)&  12.96(0.02) &  0.83(0.04) &  0.020 &  -18.524(0.124) & -18.289(0.094) & \citet{contreras18} \\
 SN\,2012ht &  N1365   &  31.879(0.043) &  13.09(0.02)  & 13.06(0.02)&  13.45(0.02)&  13.62(0.02) &  1.25(0.04) &  0.020 &  -18.516(0.113) & -18.323(0.082) & \citet{burns18}  \\
 SN\,2015F  &  N2442   &  31.482(0.053) &  13.47(0.02)  & 13.26(0.02)&  13.12(0.05)&  13.42(0.10) &  1.23(0.04) &  0.180 &  -18.606(0.125) & -18.225(0.131) & \citet{burns18}  \\
\hline
Mean               &  &              &               &            &              &             &              &        &  -18.530         &  -18.297       &                \\
$\sigma$           &  &              &               &            &              &             &              &        &    0.086         &   0.101        &                \\
$\sigma$/$\sqrt n$ &  &              &               &            &              &             &              &        &    0.029         &   0.034        &                \\
Error in Mean      &  &              &               &            &              &             &              &        &    0.043         &   0.034        &                \\
$n$                &  &              &               &            &              &             &              &        &    9             &   9            &                \\
\hline
\end{tabular}
\end{table}
\end{landscape}

\subsection{THE F19 METHODOLOGY} \label{F19}

\citet{freedman19} recently revisited the determination of \Ho\ from 99 CSP-I distant SNe using the light curve analysis developed by \citet[B18]{burns18}, in which the SN magnitudes are modeled using the light curve fitter SNooPy
\citep{burns11,burns14}, which delivers for each SN its peak magnitudes corrected for K-terms and Galactic reddening,
and $s_{BV}$ which is the colour-stretch parameter (equivalent to the decline rate \dm). As described by \citetalias{burns18}, the standardization of the SN luminosities is performed using two approaches, the ``Reddening'' and the ``Tripp'' models . 
The former has the form,

\begin{equation}
m_X = P^0 + P^1(s_{BV}-1) + P^2(s_{BV}-1)^2 +  \mu(z,H_0,C) + R_X E(B-V) + \alpha_M (log \frac{M_*}{M_\odot} - M_0)
\end{equation}

\smallskip
\noindent Similarly to \citetalias{folatelli10}, this model computes peak magnitude corrections for decline rate
(the linear and quadratic $s_{BV}$ terms), and for host galaxy reddening using the colour excess $E(B-V)$ derived
from optical and NIR colours of each SN, but incorporates an additional correction due to the SN host galaxy
stellar mass, $M_*$, obtained from the $H$-band magnitude of the host galaxy. In this equation $m_X$ is the SN
observed peak magnitude in band $X$ and $\mu(z,H_0,C)$ is the distance modulus computed from the SN host galaxy
redshift given a set of cosmological parameters \Ho=72 \kmsmpc, $\Omega_m$=0.27, and $\Omega_\Lambda$=0.73
(see Eq. (9) of \citetalias{burns18}). In this model there are five fitting parameters: the two polynomial coefficients that
describe the luminosity versus stretch dependence, $P^1$ and $P^2$, the slope of the luminosity versus colour-excess,
$R_X$, the slope of the luminosity versus host galaxy mass, $\alpha_M$, and  $P^0$ (the peak absolute magnitude
of a SN with $s_{BV}$=1, $E(B-V)$=0, $M_*$=10$^{11}M_\odot$). 
As shown by \citetalias{burns18} the ``Reddening'' approach applied to the CSP SNe yields standardized absolute magnitudes with 
characteristic dispersions ($\sigma_X$) between 0.08-0.12 mag, with the exception of the $u$ band where the scatter is $\sim$ 0.16 mag.

\smallskip
The second approach used by \citetalias{freedman19} is the ``Tripp'' model which has the form,

\begin{equation}
m_X = P^0 + P^1(s_{BV}-1) + P^2(s_{BV}-1)^2 + \mu(z,H_0,C) + R_X (B-V) + \alpha_M (log \frac{M_*}{M_\odot} - M_0)
\end{equation}

\smallskip
\noindent The main difference between the ``Tripp'' and the ``Reddening'' models is in the way the host galaxy
reddening is addressed. Here the colour excess is replaced by $B-V$, that is, the colour of the SN at peak.
In other words, the reddening correction in the ``Tripp'' approach does not require to know the intrinsic
colour of the SN, but neglects the fact that the intrinsic colour varies with decline rate. Thus, since the
$B-V$ colour is affected both by the intrinsic and dust extinction effects, the inferred value of the $R_X$
parameter cannot be directly interpreted as a dust extinction law. As shown by \citetalias{burns18} the ``Tripp'' approach applied to the CSP SNe yields standardized absolute magnitudes with 
characteristic dispersions ($\sigma_X$) between 0.11-0.13 mag with a slight decrease toward longer wavelengths, except for the $u$ band where the
scatter is significantly higher $\sim$ 0.22 mag.

\subsubsection{Absolute Magnitudes} \label{F19_AM}

\citetalias{freedman19} presented in column 6 of Table 3 standardized apparent peak magnitudes in the $B$-band
for 27 nearby SNe, using the “Tripp” model. We employ such data in order to calculate absolute peak magnitudes
using the 18 nearby SNe which have TRGB distances, from which we derive a mean value of
$M^B_{\mathrm{MAX,corr}}$=-19.223$\pm$0.029, which compares well with the -19.225$\pm$0.029 published by
\citetalias{freedman19}. Then we repeat the same procedure but this time using the 19 nearby SNe with
Cepheid distances, which yield $M^B_{\mathrm{MAX,corr}}$=-19.150$\pm$0.033, after adding a correction of
-0.029 mag (= 5 log 73.24/74.22) in order to place this value in the \citetalias{riess19} Cepheid scale.

\subsubsection{The Hubble-Lema\^itre constant} \label{F19_H0}

\citet{freedman19} applied the two approaches described by \citetalias{burns18} to a subset of 99 CSP distant SNe with $BiJHK$ light curves and meeting the requirements $E(B-V)$ \textless 0.5 and $s_{BV}$ \textgreater 0.5, and presented in Table 5 
individual \Ho\ values for the $BiJHK$ filters, using both the ``Reddening'' and the ``Tripp'' models. Here we attempt to reproduce their results but we face various problems, namely, (1) \citetalias{freedman19} published standardized apparent peak magnitudes for the nearby SNe only for the single case of the $B$-band and the ``Tripp'' model (see their Table 3), (2) \citetalias{freedman19} did not publish the zero points of the distant Hubble diagram for any of the $BiJHK$ bands. Hence, we are only able to calculate the value of \Ho\ for that single case and {\it presuming that \citetalias{freedman19} used the same zero point published by B18}, namely $P^0(B)$=-19.162 (see their Table 1). For this purpose we employ the formula,

\begin{equation}
H_0(B) = 72 \times 10^{0.2[M^B_{\mathrm{MAX,corr}}-P^0(B)]}
\label{eq:F19}
\end{equation}

\smallskip
\noindent where we compute $M^B_{\mathrm{MAX,corr}}$ using the same data published in Table 3 of \citetalias{freedman19}
($m_{B\prime}^{\mathrm{CSP}}$ and $\mu_{\mathrm{TRGB}})$ and adopt $P^0(B)$ from \citetalias{burns18}. Our result, presented in Table \ref{H0_table}, $H_0(B)$=70.0$\pm$1.0, is 0.5\% {\it higher} than the published value by \citetalias{freedman19}, namely 69.7$\pm$1.4 \kmsmpc, thus implying that \citetalias{freedman19} did not exactly use the zero point derived by \citetalias{burns18}. Given the relevance of this topic, it is important that \citetalias{freedman19} make available all the necessary ingredients required to reproduce their results.

\smallskip
We repeat the same exercise but this time adopting the Cepheid distance moduli listed in Table 3 of \citetalias{freedman19} (with the only caveat that we add a correction of -0.029 mag in order to place such values in the \citetalias{riess19} Cepheid scale), from which we obtain $H_0(B)$=72.4$\pm$1.1 (see Table \ref{H0_table}). This value can be compared to the corresponding value obtained by \citetalias{burns18} using the same ``Tripp'' model, duly modified to the \citetalias{riess19} scale, namely, $H_0(B)$=73.7$\pm$2.1 \kmsmpc. The question that arises is, what causes this 1.3 \kmsmpc\ difference? Although it may not seem statistically significant, it proves concerning considering that both used the same method for standardizing the CSP peak magnitudes, so that the difference likely originates in the the $P^0(B)$ parameter, whose error (usually less than 0.01 mag) has an impact  of less than 0.3 \kmsmpc\ in $H_0(B)$ (see equation \ref{eq:F19}).

\subsection{THE R19 METHODOLOGY} \label{R19}

\citetalias{riess16} made a determination of the Hubble-Lema\^itre constant from a sample of 19 nearby SNe with Cepheid distances
(the \citetalias{riess16} Cepheid scale), combined with a sample of 217 distant SNe\,Ia observed with optical filters
in the course of the CSP and CfA surveys. Their $u^\prime$$g^\prime$$r^\prime$$i^\prime$$UBVRI$ light curves were
re-calibrated using the ``Supercal'' method developed by \citet{scolnic15} with the purpose to place different
SN samples on a single, consistent photometric system. The resulting light curves were analyzed with the SALT2
light curve fitter model which delivers SN peak magnitudes standardized using a colour and a stretch parameter similar
to \dm.

\smallskip
Adopting this formalism, \citetalias{riess16} obtained a $B$-band Hubble diagram with a zero point of $a_B$=0.71273$\pm$0.00176.
When combined with the Cepheid distances to 19 nearby SNe obtained by the SH0ES program, R16 derived a value of
$H_0(B)$=73.24$\pm$1.74 \kmsmpc, anchored to NGC 4258, the Milky Way and the LMC. More recently, \citetalias{riess19} presented an improved determination of \Ho\ from Hubble Space Telescope (HST) observations of Cepheids in the LMC. Using only the LMC DEBs to calibrate the Cepheid luminosities, \citetalias{riess19} derived a 1.34\% greater value than
\citetalias{riess16}, namely, $H_0(B)$=74.22$\pm$1.82 \kmsmpc.

\subsubsection{Absolute Magnitudes} \label{R19_AM}

\citetalias{riess16} presented in Table 5 standardized apparent peak magnitudes in the $B$-band (column 3) for the 19 SNe
with Cepheid distances (column 5). We employ such data in order to calculate absolute peak magnitudes, from which
we derive a mean value of $M^B_{\mathrm{MAX,corr}}$=-19.251$\pm$0.036 in the \citetalias{riess16} Cepheid scale. Unfortunately,
\citetalias{riess19} did not publish the individual Cepheid distances re-calibrated to the LMC distance alone. Despite
this difficulty, we manage to add a correction of -0.029 mag (= 5 log 73.24/74.22) to the \citetalias{riess16}
distance moduli in order to place them in the \citetalias{riess19} scale, from which we derive a mean absolute magnitude
$M^B_{\mathrm{MAX,corr}}$=-19.222$\pm$0.036. Now we repeat the same procedure but this time using the subset of ten nearby
SNe with TRGB distances \citepalias{freedman19}, from which we obtain $M^B_{\mathrm{MAX,corr}}$=-19.326$\pm$0.038,
which is identical to that obtained by \citetalias{freedman19}.

\subsubsection{The Hubble-Lema\^itre constant} \label{R19_H0}

As mentioned above, \citetalias{riess19} obtained $H_0(B)$=74.22$\pm$1.82 \kmsmpc, when using solely the LMC DEBs
to calibrate the Cepheid luminosities. Here we attempt to reproduce their result using their equation 9,

\begin{equation}
log H_0(B) =  \frac {M^0_B + 5a_B + 25}{5}
\end{equation}

\smallskip
\noindent where $M^0_B$ is the mean standardized $B$ band peak magnitude -19.222$\pm$0.036 in the \citetalias{riess19}
Cepheid scale and $a_B$ is the zero point of the $B$ band Hubble diagram, 0.71273$\pm$0.00176. Our result, presented in Table \ref{H0_table}, $H_0(B)$=73.8$\pm$1.2 is 0.5\% {\it lower} than the published value by \citet{riess19}, most likely due to the fact that we do not have access to the individual \citetalias{riess19} Cepheid distances. Applying this formula to the mean magnitude
-19.326$\pm$0.038 obtained from the ten TRGB distances, we obtain $H_0(B)$=70.4$\pm$1.2 \kmsmpc.

\bsp 
\label{lastpage}
\end{document}